\documentclass[11pt,a4paper]{article}
\usepackage[T1]{fontenc}
\usepackage[utf8]{inputenc}
\usepackage{lmodern}
\usepackage[margin=25mm]{geometry}
\usepackage{amsmath,amssymb,booktabs,array}
\usepackage{microtype}
\usepackage[hidelinks]{hyperref}
\usepackage[dvipsnames]{xcolor}
\usepackage{cite}
\usepackage{multirow}
\usepackage{slashed}
\usepackage{verbatim} %
\usepackage{enumitem}
\usepackage{rotating}
\usepackage{xcolor}
\usepackage{multirow}
\usepackage{bm}
\usepackage[normalem]{ulem}
\usepackage{cleveref}
\usepackage{booktabs} 
\usepackage{tikz-cd} 
\usepackage{empheq}
\usepackage{mathrsfs}

\graphicspath{ {figures/} }

\usepackage{tikz}
\usepackage{diagbox}
\usetikzlibrary{positioning}
\usetikzlibrary{calc}
\usetikzlibrary{decorations.pathreplacing,calligraphy}

\usepackage{xstring}
\usetikzlibrary{decorations.pathmorphing} 
\usetikzlibrary{decorations.markings} 
\usetikzlibrary{arrows} 
\usetikzlibrary{shapes} 
\usetikzlibrary{matrix} 
\usetikzlibrary{positioning} 
\usepackage[english]{babel} 
\usepackage[autostyle]{csquotes}
\usepackage{pifont}
\usetikzlibrary{shapes.multipart}
\usetikzlibrary{bending}

\tikzset{->-/.style={decoration={
  markings,
  mark=at position .5 with {\arrow{>}}},postaction={decorate}}}

\newcommand{\Neu}{\text{Neu}}
\newcommand{\Dir}{\text{Dir}}
\newcommand{\Bsym}{\mathfrak{B}^{\text{sym}}}
\newcommand{\Bphys}{\mathfrak{B}^{\text{phys}}}

\renewcommand{\dim}{\text{dim}}

\newcommand{\ba}{\begin{aligned}}
\newcommand{\ea}{\end{aligned}}

\renewcommand{\d}{\mathrm{d}}

\numberwithin{equation}{section}

\newcommand{\be}{\begin{equation}} \newcommand{\ee}{\end{equation}}
\newcommand{\bea}{\begin{equation} \begin{aligned}} \newcommand{\eea}{\end{aligned} \end{equation}}

\newcommand{\cA}{\mathcal{A}}

\newcommand{\cD}{\mathcal{D}}

\newcommand{\cF}{\mathcal{F}}

\newcommand{\cH}{\mathcal{H}}
\newcommand{\cI}{\mathcal{I}}

\newcommand{\cO}{\mathcal{O}}
\newcommand{\cP}{\mathcal{P}}
\newcommand{\cQ}{\mathcal{Q}}

\newcommand{\cS}{\mathcal{S}}

\newcommand{\cZ}{\mathcal{Z}}

\newcommand{\bN}{\mathbb{N}}

\newcommand{\bQ}{\mathbb{Q}}
\newcommand{\bR}{\mathbb{R}}

\newcommand{\bZ}{\mathbb{Z}}

\newcommand{\fB}{\mathfrak{B}}

\def\repa{\raise4pt\hbox{$\square$}\mkern-14mu\raise-4pt\hbox{$\square$}}
\def\repab{\overline{\raise4pt\hbox{$\square$}\mkern-14mu\raise-4pt\hbox{$\square$}\mkern-1mu}}

\begin{document}

\baselineskip=14pt  %
\numberwithin{equation}{section}  %
\allowdisplaybreaks  %

\thispagestyle{empty}

\vspace*{0.8cm} 
\begin{center}
{{\Huge Decoupling Dynamics: A Tale of Two Limits}}

\vspace*{0.8cm}
{\large Andrea Antinucci and Oscar Lewis}

\vspace*{.4cm} 
{\it  Mathematical Institute, University of Oxford, \\
Andrew Wiles Building,  Woodstock Road, Oxford, OX2 6GG, UK}

\vspace*{0cm}
 \end{center}
\vspace*{0.4cm}

\begin{abstract}
\noindent The symmetries of a $d$-dimensional QFT are known to be captured by a topological theory in one dimension higher, the Symmetry Topological Field Theory (SymTFT), which is invariant under gauging finite symmetries of the QFT, but not under dynamical gauging of continuous symmetries. We show that the distinct SymTFTs of a $d$-dimensional QFT before and after dynamical gauging of a continuous Abelian symmetry  arise as different topological limits of the same dynamical holographic gauge theory in $(d+1)$-dimensions. Gauging changes the boundary condition of this dynamical theory and thereby selects a different topological limit, explaining why the associated SymTFT changes. We give a fully quantum definition of these SymTFTs at the level of Hilbert spaces and operators by combining suitable limits of the bulk holographic theory with photon truncation. This construction clarifies how topological symmetry defects, charges, and anomalies survive while the dynamical information carried by currents is discarded. We study non-anomalous and anomalous examples, providing a complete quantum analysis for $(2+1)$-dimensional bulks. For a four-dimensional boundary theory with a mixed axial-vector anomaly, we recover the SymTFT of the non-invertible chiral symmetry that remains after gauging the vector symmetry.
\end{abstract}

\newpage

\tableofcontents

\section{Introduction}
How much information associated with a symmetry is captured purely topologically? For discrete symmetries of a covariant QFT $T$, the symmetry structure is encoded by topological symmetry defects: their fusion, junctions and allowed deformations organize anomalies, gauging and possible patterns of spontaneous symmetry breaking \cite{Gaiotto:2014kfa}. The Symmetry Topological Field Theory (SymTFT) framework formalizes this idea through a topological quantum field theory $\cZ(\cS)$ in one dimension higher, which separates the structural data of the symmetry $\cS$ from the dynamics of $T$ \cite{Gaiotto:2014kfa, Gaiotto:2020iye, Ji:2019jhk, Apruzzi:2021nmk, Freed:2022qnc, Kaidi:2022cpf, Bhardwaj:2023ayw}. The theory $T$ is recovered by compactifying $\cZ(\cS)$ on an interval, with a physical boundary condition $\fB_{\text{phys}}(T)$ carrying the dynamics and a topological boundary condition $\fB_{\text{sym}}(\cS)$ supporting the symmetry (see Fig \ref{fig:placeholder}). Gauging an anomaly free sub-symmetry of $\cS$ amounts to replacing the latter by another topological boundary condition $\fB_{\text{sym}}'$, and is itself a topological operation.

The framework also extends to continuous symmetries \cite{Antinucci:2024zjp, Brennan:2024fgj, Bonetti:2024cjk, Antinucci:2024bcm} including space-time ones \cite{Apruzzi:2025hvs}.\footnote{One should notice that the TQFTs used in this context typically have infinite partition function and therefore don't satisfy all the axioms of a TQFT \cite{Atiyah:1989vu}. However, they are topological theories in the physical sense of having a vanishing stress energy tensor, and mathematically they are formalized as once-categorified TQFTs \cite{Freed:2009qp, Stockall:2025ngz}. See also \cite{Cattaneo:1997eh, Cattaneo:2000rt, Cattaneo:2000mc, Cattaneo:2002tk} for previous studies, and \cite{Jia:2025vrj, Guan:2026zsk} for recent developments.} From a physical viewpoint, an important qualification that distinguishes the scope of the SymTFTs for finite and continuous symmetries is that when dealing with continuous symmetries some symmetry-related data is dynamical.
Continuous symmetries typically admit gauge-invariant current operators\footnote{Exceptions occur in non-compact gauge theories. Two identical non-compact Maxwell fields, for example, admit an $SO(2)$ symmetry rotating them. The associated Noether current is not gauge invariant, although its global charge is gauge invariant when boundary terms vanish.}, whose correlation functions depend on more than topology and are not fixed by the symmetry alone. The continuous SymTFT encodes coupling to flat backgrounds and topological information such as anomalies, but it should not capture the full dynamical response to general non-flat backgrounds which probes current correlation functions. Consequently, changing its topological symmetry boundary describes gauging discrete sub-symmetries or flat gauging the full symmetry, but cannot implement ordinary \emph{dynamical gauging}, where the gauge field becomes fully dynamical with non-trivial field strength.
\begin{figure}[t]
    \centering
\begin{tikzpicture}
\fill[blue!20] (0,0) rectangle (6,4);
\draw[thick] (0,0) -- (0,4);
\draw[thick] (6,0) -- (6,4);
\filldraw[black] (6,-0.2) circle (0pt) node[anchor=north]{\large{$\Bphys(T)$}};
\filldraw[black] (0,-0.2) circle (0pt) node[anchor=north]{\large{$\Bsym(\cS)$}};
\filldraw[black] (3,2) circle (0pt) node[]{\large{$\cZ(\cS)$}};
\end{tikzpicture}
\caption{The bulk TQFT $\cZ(\cS)$ on an interval with two boundary conditions: $\Bphys$ supporting the local dynamics of $T$, and $\Bsym$ supporting the symmetry $\cS$. Shrinking the interval reproduces the $d-$dimensional theory $T$ with symmetry $\cS$.}
    \label{fig:placeholder}
\end{figure}

Nevertheless, \cite{Antinucci:2024zjp} proposed that the dynamical gauging of a $U(1)$ symmetry in a $d$-dimensional theory $T$ induces a map between two different $(d+1)$-dimensional SymTFTs, $\cZ(\cS)\rightarrow\cZ(\cS')$, with $\cS'$ the symmetry of the gauged theory $T'$. For example, if $\cS$ is a non-anomalous $U(1)$ $p$-form symmetry in $d$ spacetime dimensions, the SymTFT has action
\begin{equation}\label{eq:basicSymTFT}
    \cZ(\cS)\,:\qquad
    \frac{i}{2\pi}\int_{M_{d+1}}b_{d-p-1}\wedge\d A_{p+1} \, ,
\end{equation}
where $b_{d-p-1}$ is a non-compact Abelian $(d-p-1)$-form gauge field and $A_{p+1}$ a compact Abelian $(p+1)$-form gauge field. The Dirichlet symmetry boundary condition fixes $A_{p+1}\big|_\partial=0$. If one dynamically gauges the $U(1)$ symmetry in the boundary theory $T$, the symmetry $\cS'$ of the gauged theory $T'$ is a non-anomalous $U(1)$ $(d-p-3)$-form symmetry. Its SymTFT is
\begin{equation}\label{eq: intro gauged SymTFT}
    \cZ(\cS')\,:\qquad
    \frac{i}{2\pi}\int_{M_{d+1}}\d \widetilde{A}_{d-p-2}\wedge \widetilde{b}_{p+2}\, ,
\end{equation}
with Dirichlet boundary conditions for $\widetilde{A}_{d-p-2}$, that describes the magnetic symmetry $\cS'=U(1)^{(d-p-3)}$ of $T'$.
Here $\widetilde{A}_{d-p-2}$ is a new compact $(d-p-2)$-form gauge field and $\widetilde{b}_{p+2}$ a new non-compact $(p+2)$-form gauge field. This \emph{dynamical gauging map} $\cZ(\cS)\rightarrow \cZ(\cS')$ is then summarized by the replacement
\begin{equation}\label{eq: intro gauging substitution}
    b_{d-p-1}\mapsto\d \widetilde{A}_{d-p-2}\, , \qquad
    \d A_{p+1}\mapsto \widetilde{b}_{p+2} \, .
\end{equation}
Let us emphasize that this map does \emph{not}, in any sense, correspond to dynamically gauging a symmetry in the bulk $(d+1)$-dimensional SymTFT, and is merely a map between two such SymTFTs that reproduces the gauging on the boundary.

In \cite{Antinucci:2024zjp} this map was motivated by coupling the matter and photon SymTFTs  and analyzing which operators can end. However, this argument was limited to the simplest case and did not explain why such a map between different TQFTs exists. Moreover, the idea of changing the SymTFT when dynamically gauging appears to be in slight tension with holography, where gauging simply changes the boundary conditions, even for continuous symmetries \cite{Witten:2003ya, Marolf:2006nd, DeWolfe:2020uzb}. The goal of this paper is to provide this missing explanation and a systematic way to apply it in the presence of additional couplings. In doing so, we give a quantum construction of the relevant continuous SymTFTs and clarify their relation to holography. Indeed for theories admitting a holographic dual, while it is well known that the SymTFT for finite symmetries  is obtained from holography by keeping only the topological terms\cite{Witten:1998wy, Apruzzi:2021nmk, Apruzzi:2021phx, GarciaEtxebarria:2022vzq, Antinucci:2022vyk, Antinucci:2022cdi, Apruzzi:2022rei, vanBeest:2022fss, Argurio:2024oym}, the relation between the continuous SymTFT and holography remained less clear.\footnote{For earlier discussions of this relation, see \cite{Bonetti:2024cjk,Mignosa:2025cpg,Bergman:2025isp}.} The input of our discussion comes from the holographic gauge sector, which retains the current dynamics through a Maxwell-type action \cite{Apruzzi:2024htg}. Dynamically gauging the boundary $U(1)$ symmetry then corresponds to changing the bulk gauge field's boundary condition from Dirichlet to Neumann. Our main results are: 
\begin{itemize}
    \item The SymTFT for a given continuous symmetry is extracted from the Maxwell-type holographic theory in two steps. The first is the zero-coupling limit, already used in classical constructions \cite{Bonetti:2024cjk,Apruzzi:2025hvs},\footnote{Related connections between gravitational dynamics and topological theories appear in \cite{Smolin:2003qu,Freidel:2005ak,MacDowell:1977jt}.} which excludes nonzero magnetic flux from the finite-energy spectrum and makes the electric global states degenerate. This is not enough\footnote{The only exception is for $d=1$, when the bulk theory is two-dimensional \cite{Witten:1992xu}. Physically the reason is that gauge fields do not propagate in 2d, where pure gauge theories are quasi-topological, and become topological in the weak-coupling limit.}: photon excitations remain, with energies independent of the coupling. We define the second step as restriction to the photon vacuum and to the operators preserving it, producing a topological theory. This operation would not produce a local QFT at generic coupling, but it does after the zero coupling limit.
    \item For the non-anomalous single-$U(1)$ theory, the image of the dynamical gauging map is obtained from the opposite, infinite-coupling limit, followed by the same photon decoupling. The electric and magnetic global sectors exchange roles. The Neumann boundary of the dynamical theory becomes Dirichlet for the compact dual gauge field, giving the SymTFT of the gauged theory.
    \item We apply the same ideas to a few other examples that include additional terms in the SymTFT representing pure or mixed anomalies. In particular, we consider two $U(1)$ symmetries with a mixed anomaly, whose holographic theory includes two Maxwell theories with electric couplings $e_1,e_2$, coupled through a mixed Chern-Simons term. The SymTFT is obtained in the $e_1,e_2\rightarrow 0$ topological corner, while the corner $e_1\rightarrow0 $, $e_2\rightarrow \infty$ gives a SymTFT for the theory where one $U(1)$ has been gauged. The spectrum of the dynamical theory as a function of the couplings is distinguished from a pure anomaly, which obstructs a smooth local strong-coupling limit, from a mixed anomaly, for which a double scaling limit with $e_1e_2$ fixed is possible. We derive the finite symmetry left by the ABJ anomaly in $d=2$, and recover the action describing the non-invertible rational-angle chiral symmetry in $d=4$, without assuming a canonical quantization of the interacting five-dimensional theory.
   
\end{itemize}

For the non-anomalous theory, the two extractions can be summarized as
\be\label{eq: intro quantum extractions}
\ba
    (\text{Maxwell},\Dir_{A_{p+1}})
    &\xrightarrow[\text{photon truncation}]{e\rightarrow0}
    (\cZ(\cS),\Dir_{A_{p+1}})\, , \cr
    (\text{Maxwell},\Neu_{A_{p+1}})
    &\xrightarrow[\text{photon truncation}]{e\rightarrow\infty}
    (\cZ(\cS'),\Dir_{\widetilde{A}_{d-p-2}}) \, .
\ea
\ee
Here the change of boundary condition implements gauging in the dynamical theory, while the two extractions give the distinct topological descriptions related by \eqref{eq: intro gauging substitution}.

The motivation for the two limits is a consequence of the definition of the SymTFT as a tool that isolates the universal symmetry based properties from the non-universal dynamical information. Here the residual dynamical information, at generic values of $e$, concerns the current operator. Its existence follows from that of the monodromy defects \cite{Gabai2026}: co-dimension $p+2$ objects on the boundary obtained by cutting open the symmetry defect. For $p=0$ monodromy defects create states whose energy cost is proportional to $e^{-2}$, motivating the weak-coupling extraction as the limit that forgets this current response. The dual argument selects strong coupling after gauging. What is discarded is the dynamical, non-universal aspect of the symmetry, while topological symmetry defects and their data survive. %

The rest of the paper develops this construction explicitly. We begin with canonical quantization in Section \ref{sec:CQ}, separating the photon and global sectors, then construct the topological subsectors and their boundary dictionary in Sections \ref{sec: topological subsector}--\ref{sec: boundary perspective}. Section \ref{sec: gauging from holography} derives the gauging map, and the applications illustrate the role of additional couplings. Finally, the higher-dimensional construction identifies the $(d+2)$-dimensional anomaly-polynomial TFT as the SymTFT of the $(d+1)$-dimensional Maxwell theory itself. Maxwell is a dynamical boundary of that further TFT, while the original theory remains $d$-dimensional.

\section{Canonical Quantization} \label{sec:CQ}

\subsection{Maxwell Theory in $(d+1)$-dimensions} \label{sec: CQ of Maxwell}

Consider Maxwell theory on $M_{d+1}=\bR \times \Sigma_d $, where $\Sigma_d$ denotes a closed, connected, oriented Riemannian manifold with metric $h$:
\be \label{eq: Maxwell action}
    S_{\text{Max}} = -\frac{1}{2 e^2} \int_{M_{d+1}} \d A_1 \wedge \star_{d+1} \d A_1 \, .
\ee
We assume throughout that $\Sigma_d$ contains no torsion in its homology or cohomology, meaning fluxes commute \cite{Freed:2006ya,Freed:2006yc}. Let us review the canonical quantization of this theory\cite{Gupta:1949rh,Bleuler1950,Henneaux:1992ig,Verlinde:1995mz,DiracLectures,Beasley:2014jta}, emphasizing the global aspects that will be crucial for us.

The components of the physical electric and magnetic fields are defined by $B_{ij}=F_{ij}$ and ${E_{\text{p}}}_i=F_{0i}$, where the subscript $\text{p}$ denotes physical. %
Rescaling the electric field $E_i := \frac{1}{e^2} {E_{\text{p}}}_i$, the momentum conjugate to the spatial components of $A_1$ is 
\be
    \Pi^i = \sqrt{h} \, h^{ij} E_j \, .
\ee
The Maxwell Hamiltonian reads
\be \label{eq: Maxwell Hamiltonian}
    H_{\text{Max}} = \frac{1}{2} \int_{\Sigma_d} \d^dx \, \sqrt{h} \, \left( e^2 E_i E_j h^{ij}+ \frac{1}{2 e^2} B_{ij} B_{kl}h^{ik}h^{jl} -2A_0 (\nabla_i E_j h^{ij}) \right) \, .
\ee
$A_0$ appears as a Lagrange multiplier that imposes the Gauss law constraint on $E$. 

It is convenient to pass to the differential form description of the Hamiltonian. Define a $U(1)$-connection $A:=A_1|_{\Sigma_d}$ on $\Sigma_d$, and $\star :=\star_d : \Omega^p(\Sigma_d,\bR) \rightarrow \Omega^{d-p}(\Sigma_d,\bR)$ is the Hodge operator on $\Sigma_d$ with respect to the metric $h$. The magnetic field is $B=F|_{\Sigma_d} \in \Omega^2(\Sigma_d,\bR)$, while we build a 1-form $E:= E_i|_{\Sigma} \, \d x^i\in \Omega^1(\Sigma_d, \bR)$. %
The Gauss law constraint becomes $\nabla_i E_j h^{ij}=-\d^* E=0$, where we introduced $\d^*|_{\Omega^p}=(-1)^{d(p-1)-1}\star \d \star$. The Hamiltonian takes the form 
\be \label{eq: Hamiltonian of Maxwell}
    H_{\text{Max}} = \frac{1}{2} \int_{\Sigma_d} \left( e^2 E\wedge \star E+\frac{1}{e^2} B \wedge \star B +2 A_0  \star (\d^* E) \right) \, .
\ee
One extracts the symplectic form:
\be \label{eq: symplectic form of Maxwell}
    \Omega_{\text{Max}}(\delta A, \delta E) = \int_{\Sigma_d} \star \delta  E \wedge \delta A \, .
\ee

To take into account the global aspects arising from the non-trivial topology of $\Sigma_d$, we expand $A$ and $E$ using the orthogonal Hodge decomposition on $(\Sigma_d,h)$ 
\be \label{eq: hodge decomposition for Maxwell}
    A = \d \lambda_0 + \star \d \sigma_{d-2} + a_{\text{h}} \, , \qquad E = \d \psi_0 + \star \d \chi_{d-2} +e_{\text{h}} \, ,
\ee
Here $\lambda_0,\psi_0 \in \Omega^0(\Sigma_d,\bR)$, $\sigma_{d-2},\chi_{d-2}\in \Omega^{d-2}(\Sigma_d,\bR)$, while $a_{\text{h}},e_{\text{h}} \in \cH^1(\Sigma_d)$ are harmonic 1-forms on $\Sigma_d$, i.e. closed and co-closed. Note that the Hodge decomposition of $A$ is only valid locally, and it is performed in a fixed magnetic flux sector. Generic magnetic flux data will be contained in the harmonic sector of $B$.

Firstly, we use the Coulomb gauge fixing condition $\d^* A = 0$ to enforce $\d \lambda_0=0$. %
Similarly the Gauss law constraint $\d^* E=0$ enforces $\d \psi_0=0$. Thus 1-forms receive no contribution from the exact sector. We will comment later how to take care of the large gauge transformations of $A$. 

Secondly, as locally we have $B= \d A$, the contribution to the magnetic field from the co-exact sector of $A$ is $\d \star \d \sigma_{d-2}$. However, this does not account for the global topology of $\Sigma_d$. Indeed, we can have that along non-trivial cycles in $H_2(M_{d+1},\bZ)$, fluxes of the field strength $F_2$ over these cycles are quantized \cite{Witten:1995gf,Verlinde:1995mz}. Using $H_2(M_{d+1},\bZ) \cong H_2(\Sigma_d,\bZ)\cong \bZ^{b_2(\Sigma_d)}$, we capture this non-trivial flux over integral 2-cycles $\Gamma^I \in H_2(\Sigma_d,\bZ)$ using the magnetic field $B$, yielding non-trivial flux sectors
\be
    \int_{\Gamma^I} \frac{B}{2\pi} = m^I \in \bZ \, .
\ee
Thus the Hodge decomposition of the magnetic field $B$ is
\be
    B = \d \star \d \sigma_{d-2} + b_{\text{h}} = \d \star \d \sigma_{d-2} + 2\pi \sum_{I=1}^{b_2(\Sigma_d)} m^I {v}_I \, .
\ee
Here we expanded $b_{\text{h}}\in \cH^2(\Sigma_d)$ onto a basis of harmonic representatives $\{ v_I \}_{I=1}^{b_2(\Sigma_d)}$ normalized such that 
\be
    \int_{\Gamma^I} v_J = {\delta^I}_J  \, , \qquad   \int_{\Sigma_d} v_I \wedge \star v_J = G_{IJ} \, .
\ee

Physically, the Maxwell Hilbert space on $\Sigma_d$ contains the "local" sector given by the ordinary photon excitations, and these come from quantizing the co-exact sector (that has in fact $d-1$ polarization degrees of freedom\footnote{To count the polarizations, observe that $\sigma_{d-2}$ has ${d \choose d-2}$ degrees of freedom. However, it is only defined up to shifts by exact forms, meaning we can shift $\sigma_{d-2} \rightarrow \sigma_{d-2}+\d \sigma_{d-3}$ for some $\sigma_{d-3}\in \Omega^{d-3}(\Sigma_d,\bR)$. We can then shift $\sigma_{d-3} \rightarrow \sigma_{d-3}+\d \sigma_{d-4}$, and so on. Taking this redundancy into account, we see there are exactly
\be 
    \sum_{n=2}^d {d \choose d-n} (-1)^n =d-1    \, ,
\ee
degrees of freedom in the co-exact sector.}), and a "global" sector coming from the harmonic sector. The excitations of the latter are created by Wilson and 't Hooft operators wrapping non-trivial cycles\footnote{Notice that a Wilson line on a contractible cycle creates a state that can be expresses as a complicated linear combination of multi-photon states. Hence only the topologically non-trivial defects create genuinely new states.}. We will be mostly interested in the global sector and how it interpolates to the topological "limits", i.e. the continuous SymTFTs. The reason the word "limits" is in quotes it is that, in fact, there is no limit in which the local sector goes away, and we need to carefully clarify what operation allows to extract a TQFT from Maxwell theory. 

So let us first deal with the co-exact sector. This is very standard. We  expand $\sigma_{d-2}$ and $\chi_{d-2}$ onto a basis $\{ \varphi_n \}$ of co-closed $(d-2)$-forms $\varphi_n \in \Omega^{d-2}(\Sigma_d,\bR)$ satisfying 
\be \label{eq: modes of laplacian Max}
    \d^* \varphi_n=0 \, , \qquad \Delta \varphi_n = \mu_n \varphi_n \, , \quad \mu_n > 0 \, , \qquad \qquad \int_{\Sigma_d} \varphi_n\wedge \star \varphi_m = \delta_{nm} \, ,
\ee
with $\Delta=\d \d^*+\d^* \d$. The mode expansions of the coexact sector then take the form
\be  \label{eq: laplacian expansion Max}
    \sigma_{d-2} = \sum_{n \in \mathbb{N}} \frac{1}{\sqrt{\mu_n}} q^n \varphi_n \, , \qquad \chi_{d-2} = \sum_{n,m \in \mathbb{N}} \frac{1}{\sqrt{\mu_n}} \delta^{nm} p_m \varphi_n \, .
\ee
Quantization promotes the position and momenta to operators with canonical quantization relation determined by \eqref{eq: symplectic form of Maxwell}:
\be \label{eq: coexact oscillator algebra}
    [\widehat{q}^n,\widehat{p}_m] = i {\delta^n}_{m} \, , \qquad [\widehat{q}^n,\widehat{q}^m]=[\widehat{p}_n,\widehat{p}_m]=0 \, .
\ee
By plugging these expansion into the Hamiltonian \eqref{eq: Hamiltonian of Maxwell} we obtain the contribution due to the co-exact sector
\be
    \widehat{H}^{\text{co}}_{\text{Max}} = \sum_{n \in \bN} \sqrt{\mu_n} \left( \widehat{a}^{\dagger}_n \widehat{a}_n + \frac{1}{2} \right) \, ,
\ee
that is the familiar photon Hamiltonian in terms of creation and annilhilation operators:
\be \label{eq: Creation annihilation ops}
    \widehat{a}_n = \frac{1}{\sqrt{2}} \left( \mu_n^{\frac{1}{4}} e^{-1} \delta_{nm}\widehat{q}^m+i e \mu_n^{-\frac{1}{4}} \widehat{p}_n \right) \, , \qquad  \widehat{a}_n^{\dagger} = \frac{1}{\sqrt{2}} \left( \mu_n^{\frac{1}{4}}e^{-1} \delta_{nm}\widehat{q}^m-i e \mu_n^{-\frac{1}{4}} \widehat{p}_n \right) \, .
\ee
This is the contribution from the photon modes, and we will refer to it as the \emph{local sector}. Importantly this is independent of the coupling $e$.

Let us now consider the global sector. 
We expand
\be \label{eq: Maxwell Harmonic form decomposition}
    a_{\text{h}} = \sum_{I=1}^{b_1(\Sigma_d)} Q^I  u_{I} \, , \qquad  e_{\text{h}}= \sum_{I,J=1}^{b_1(\Sigma_d)} L^{IJ} P_J u_I \, ,
\ee
using a basis $\{ u_I \}_{I=1}^{b_1(\Sigma_d)}$ of $H^1(\Sigma_d,\bZ)$, normalized as
\be \label{eq: orthonormal basis of harmonic forms}
    \int_{\gamma^I} u_J = {\delta^I}_J \, , \qquad  \int_{\Sigma_d} u_I \wedge \star u_J =L_{IJ} \, .
\ee
Here $\gamma^I \in H_1(\Sigma_d,\bZ)$ is the integral cycle dual to $u_I$. 

At this point we need to quotient by large gauge transformations of $A$. While harmonic forms are in $\cH^1(\Sigma_d)\cong H^1(\Sigma_d, \bR)$, we must further identify states related by elements of $H^1(\Sigma_d, \bZ)\subset H^1(\Sigma_d, \bR)$. This is precisely the redundancy of large gauge transformations, and forces the quotient $H^1(\Sigma_d, \bR)/2\pi H^1(\Sigma_d,\bZ)\cong H^1(\Sigma_d, U(1))$. Given that $u_I \in H^1(\Sigma_d,\bZ)$, this quotient amounts to the identification $Q^I \sim Q^I+2\pi n^I$, $n^I \in \bZ$. By quantizing, \eqref{eq: symplectic form of Maxwell} implies the algebra 
\be \label{eq: canonical comm relations harmonic sector}
    [\widehat{Q}^I , \widehat{P}_J] = i {\delta^I}_{J} \, , \qquad [\widehat{Q}^I,\widehat{Q}^J]=[\widehat{P}_I,\widehat{P}_J] = 0 \, ,
\ee
and the periodicity of $\widehat{Q}^I$ implies that $\widehat{P}_I$ have discrete spectrum.
 
 One further quantizes the flux sector generators $\widehat{m}^I$ appearing in the magnetic field harmonic sector, with commutation relations
 \be
    [\widehat{m}^I,\widehat{m}^J]=[\widehat{m}^I,\widehat{Q}^J]=[\widehat{m}^I,\widehat{P}_J]=0 \, .
 \ee
 Inserting these decompositions into the Hamiltonian \eqref{eq: Hamiltonian of Maxwell}, we acquire the total harmonic sector contribution
\be
    \widehat{H}^{\text{harm}}_{\text{Max}} = \frac{e^2}{2} \sum_{I,J=1}^{b_1(\Sigma_d)} L^{IJ} \widehat{P}_I \widehat{P}_J  + \frac{2\pi^2}{e^2} \sum_{I,J=1}^{b_2(\Sigma_d)} G_{IJ} \widehat{m}^I \widehat{m}^J    \, ,
\ee
where $L^{IJ}$ denotes the inverse of $L_{IJ}$. 
Putting all of this together, we get the full normal ordered Hamiltonian of Maxwell theory
\be
\mathopen{:}\widehat{H}_{\text{Max}}\mathclose{:}
    = \sum_{n \in \bN} \sqrt{\mu_n} \widehat{a}^{\dagger}_n \widehat{a}_n + \frac{e^2}{2} \sum_{I,J=1}^{b_1(\Sigma_d)} L^{IJ} \widehat{P}_I \widehat{P}_J  + \frac{2\pi^2}{e^2} \sum_{I,J=1}^{b_2(\Sigma_d)} G_{IJ} \widehat{m}^I \widehat{m}^J \, .
\ee

Lastly, let us comment on the structure of the phase space and Hilbert space. The phase space of the harmonic sector with zero magnetic flux can be identified with the cotangent bundle $T^*H^1(\Sigma_d,U(1))\cong T^*U(1)^{b_1(\Sigma_d)}$, with Darboux coordinates $(Q^I,P_I)$. The co-exact sector contributes the structure of an infinite dimensional cotangent bundle, equipped with canonical coordinates $(q^n,p_n)$. Combining these, and summing over magnetic flux sectors $\mathbf{m}=(m^I)$, we get the full Maxwell theory phase space. 
The Hilbert space can then be defined as $L^2$ functions on configuration space, the base of the phase bundle. In this case, we get $L^2(H^1(\Sigma, U(1)))$ from the zero flux harmonic sector contribution. The contribution from the co-exact sector is the usual Fock space $\mathcal{F}$, giving the total Hilbert space of Maxwell theory
\be
    \cH_{\text{Max}} = \bigoplus_{\mathbf{m} \in \bZ^{b_2(\Sigma_d)}} \left\{ L^2(H^1(\Sigma_d,U(1)))  \otimes \cF \right\}_{\mathbf{m}} \cong \bigoplus_{\substack{m \in \bZ^{b_2(\Sigma_d)} \\ n \in \bZ^{b_1(\Sigma_d)}}} \mathcal{F}   \, .
\ee
The second presentation of the Hilbert space makes electric-magnetic duality manifest. We label states in this Hilbert space
\be
\left| \{ N_r \}_{r\in \bN} ; \{ n_I \}_{I=1}^{b_1(\Sigma_{d})}, \{ m^I \}_{I=1}^{b_2(\Sigma_d)} \right\rangle \, ,
\ee
where $\{ N_r \}$ are the photon occupation numbers, $\{ n_I \}$ label the Wilson lines wrapped on non-trivial integral cycles, and $\{ m^I \}$ label the magnetic flux through integral 2-cycles.

\subsection{BF Theory in $(d+1)$-dimensions} \label{sec: CQ of BF}

Quantization of BF theory \cite{Horowitz:1989ng,Blau:1989bq,Birmingham:1991ty,Bergeron:1994ym,Szabo:1999gm, Cattaneo:1997eh, Cattaneo:2000mc, Cattaneo:2000rt} works in much the same way, and we present it briefly. The $U(1)$ BF theory has action
\be
    S_{\text{BF}} = -\frac{1}{2 \pi} \int_{M_{d+1}} b_{d-1} \wedge \d A_1 \, ,
\ee
where $A_1$ is a $U(1)$ gauge field and $b_{d-1}$ an $\bR$ $(d-1)$-form gauge field. The action has local gauge redundancies for both $A_1$ and $b_{d-1}$. There are no large gauge transformations for $b_{d-1}$, as it is globally defined. The momentum conjugate to $A_1$ is 
\be
    \Pi^i = -\frac{1}{2\pi (d-1)!} \varepsilon^{i i_1 \cdots i_{d-1}} b_{i_1 \cdots i_{d-1}} \, . 
\ee
The momentum conjugate to $b_{d-1}$ vanishes. The Hamiltonian on $\Sigma_d$ is
\be \label{eq: Hamiltonian of BF}
    H_{\text{BF}} = \frac{1}{2\pi} \int_{\Sigma_d} (b_0 \wedge \d A + A_0 \d b) \, ,
\ee
where $A_1=A+A_0 \d t$ and $b_{d-1}=b +\d t \wedge b_0$.  $A_0$ and $b_0$ are Lagrange multipliers imposing $\d A=\d b=0$, after which the Hamiltonian is identically zero. The symplectic form is
\be \label{eq: symplectic form of BF}
    \Omega_{\text{BF}}(\delta A, \delta b) = \frac{(-1)^d}{2\pi} \int_{\Sigma_d} \delta b \wedge \delta A \, .
\ee

Using Hodge decomposition on $\Sigma_d$
\be \begin{array}{cc}
A = \d \lambda_0 + \star \d \sigma_{d-2} + a_{\text{h}}     & \qquad \qquad   \lambda_0,\chi_0 \in \Omega^0(\Sigma_d,\bR) \ , \ \ a_{\text{h}}\in \cH^{1}(\Sigma_d) \ \ \ \  \vspace{0.3cm} \\
 b=\d \psi_{d-2} +\star \d \chi_0 + b_{\text{h}}    & \qquad \qquad \sigma_{d-2},\psi_{d-2}\in \Omega^{d-2}(\Sigma_d,\bR) \ , \ \ b_{\text{h}}\in \cH^{d-1}(\Sigma_d) \ ,
\end{array}
\ee
the constraints $\d A=0$, $\d b=0$ eliminates the co-exact sector, and imposing the Coulomb gauge fixing conditions $\d^*A=0$, $\d^*b=0$ eliminates the exact sector. Hence the BF gauge fields are purely harmonic. We expand 
\be
    A = \sum_{I=1}^{b_1(\Sigma_d)} Q^I u_I \, , \qquad b = -2\pi \sum_{I=1}^{b_{d-1}(\Sigma_d)} (\cI^{-1})^{IJ} P_J v_I \, ,
\ee
where $\{ u_I \}_{I=1}^{b_1(\Sigma_d)}$, $\{ v_I \}_{I=1}^{b_{d-1}(\Sigma_d)}$ are harmonic bases of $H^1(\Sigma_d,\bZ)$ and $H^{d-1}(\Sigma_d,\bZ)$ normalized as
\be
    \int_{\gamma^I} u_J = {\delta^I}_J \, , \qquad \int_{\Gamma^I} v_J = {\delta^I}_J \, , \qquad \int_{\Sigma_d} u_I \wedge v_J = \mathcal{I}_{IJ} \,,
\ee
with $\gamma^I\in H^1(\Sigma_d,\bZ)$ and $\Gamma^I \in H_{d-1}(\Sigma_d,\bZ)$ the integral cycles dual to $u_I$ and $v_I$ respectively. $\mathcal{I}$ is the usual unimodular intersection matrix that follows from Poincare duality. Quotienting by large gauge transformations of $A$ imposes that $A \in H^1(\Sigma_d,U(1))$. Inverting the Poisson brackets from the symplectic form, one gets the algebra of modes \eqref{eq: canonical comm relations harmonic sector} for $\widehat{Q}^I$ and $\widehat{P}_J$. 

The classical phase space is $\cP_{\text{BF}}=H^1(\Sigma_d,U(1))\times H^{d-1}(\Sigma_d,\bR)\cong T^*(H^1(\Sigma_d,U(1))$. Therefore the Hilbert space is 
\be
    \cH_{\text{BF}} = L^2(H^1(\Sigma_d,U(1))) \, .
\ee

\paragraph{Higher-form case.} Similarly, one can treat the more general $(p+1)$-form BF theory
\be
    S_{(p+1)\text{-BF}} = -\frac{1}{2\pi} \int_{M_{d+1}} b_{d-p-1}  \wedge  \d A_{p+1} \, .
\ee
The story is completely analogous as above, so we present here only the result to fix the notation. The mode expansions are
\be
    A = \sum_{I=1}^{b_{p+1}(\Sigma_d)} Q^I u_I \,, \qquad b=-2\pi (-1)^{p(d-p-1)}\sum_{I=1}^{b_{d-p-1}(\Sigma_d)} (\cI_{p+1}^{-1})^{IJ} P_J v_I \, .
\ee
in terms of the harmonic bases $\{ u_I \}_{I=1}^{b_{p+1}(\Sigma_d)}$, $\{ v_I \}_{I=1}^{b_{d-p-1}(\Sigma_d)}$ of $H^{p+1}(\Sigma_d,\bZ)$ and $H^{d-p-1}(\Sigma_d,\bZ)$ normalized as before. The canonical commutation relations for the modes $\widehat{Q}^I$ and $\widehat{P}_J$ again satisfy \eqref{eq: canonical comm relations harmonic sector}, and the Hilbert space structure is
\be
    \cH_{(p+1)\text{-BF}} = L^2(H^{p+1}(\Sigma_d,U(1))) \, .
\ee

\subsection{3d Maxwell-Chern-Simons and BF-Chern-Simons} \label{sec: MCS and BFCS CQ}

We will also discuss anomalous symmetries, that are described holographically by the addition of Chern-Simons terms. Let us then review the quantization of simplest example, namely Maxwell-Chern-Simons theory (MCS) in $(2+1)$-dimensions \cite{SCHONFELD1981157,Deser:1982vy,Dunne:1998qy,Devecchi:1994ha}:
\be
    S_{\text{MCS}}= -\frac{1}{2e^2}\int_{M_{3}} \d A_1 \wedge \star_{3} \d A_1 - \frac{k}{4\pi} \int_{M_3} A_1 \wedge \d A_1 \, .
\ee
We assume that $k\in 2\bZ$ so that the theory is bosonic. This theory has no genuine gauge invariant monopole operators. Moreover, the $U(1)^{(1)}$ electric 1-form symmetry of pure Maxwell is broken down to $\bZ_k^{(1)}$. The topological operators for it are
\be
    U_{n}[\gamma] := \exp \left( \frac{2\pi in}{k}\oint_{\gamma} \frac{\star_3 F_2}{e^2} + in \oint_{\gamma} A_1 \right) \, , \ \ \  \ \ \ \ n \in \bZ_k
\ee

We quantize on a closed spatial surface $\Sigma_2$ of genus $g$. The Chern-Simons term modifies the electric Gauss law
\be
    -\d \star E +\frac{k}{2\pi}B = 0 \, .
\ee
Importantly, integrating this over $\Sigma_2$ imposes the constraint $km=0$, where $m$ denotes the total magnetic flux through the surface. This removes all nonzero magnetic-flux sectors. 

Following the Hodge decomposition \eqref{eq: hodge decomposition for Maxwell}, the Gauss law removes the exact electric sector as an independent degree of freedom, implying $\psi_0=\frac{k}{2\pi}\sigma_0$. The mode decomposition is almost identical to the previous section. The main differences are that the Hamiltonian receives contributions from the non-vanishing exact sector of the electric field, while there is no contribution from the harmonic sector of $B$. The Hamiltonian reads:
\be \label{eq: MCS Hamiltonian}
\mathopen{:}\widehat{H}_{\text{MCS}}\mathclose{:}
    = \sum_{n \in \bN} \sqrt{\mu_n+\frac{k^2 e^4}{4\pi^2}} \widehat{a}^{\dagger}_n \widehat{a}_n + \frac{e^2}{2} \sum_{I,J=1}^{b_1(\Sigma_2)} L^{IJ} \widehat{P}_I \widehat{P}_J -\frac{g|k|e^2}{4\pi}  \, .
\ee
Here the creation and annihilation operators \eqref{eq: Creation annihilation ops} are 
\be
\ba
    \widehat{a}_n &= \frac{1}{\sqrt{2}} \left( \left(\mu_n +\frac{k^2 e^4}{4\pi^2}  \right)^{\frac{1}{4}} e^{-1} \delta_{nm}\widehat{q}^m+i e \left(\mu_n +\frac{k^2 e^4}{4\pi^2}  \right)^{-\frac{1}{4}} \widehat{p}_n \right) \cr
    \widehat{a}_n^{\dagger} &= \frac{1}{\sqrt{2}} \left( \left(\mu_n +\frac{k^2 e^4}{4\pi^2}  \right)^{\frac{1}{4}}e^{-1} \delta_{nm}\widehat{q}^m-i e \left(\mu_n +\frac{k^2 e^4}{4\pi^2}  \right)^{-\frac{1}{4}} \widehat{p}_n \right) \, ,
\ea
\ee
while $\widehat{q}_n,\widehat{p}_n$ are defined as before. The modes for exact and coexact sector are unchanged and still satsify \eqref{eq: coexact oscillator algebra}. The important difference is the algebra for the harmonic sector\footnote{Since $L^{IJ}$ is symmetric $\widehat{H}_{\text{MCS}}$ does not suffer from a Weyl-ordering ambiguity.}:
\be \label{eq: Harmonic mode algebra with K}
    [\widehat{Q}^I,\widehat{Q}^J] = 0 \, , \qquad [\widehat{Q}^I,\widehat{P}_J] = i {\delta^I}_J \, , \qquad [\widehat{P}_I,\widehat{P}_J] = \frac{ik}{2\pi} \cI_{IJ} \, .
\ee
Unlike in pure Maxwell theory there is a zero-point energy removed from the harmonic sector by normal ordering, arising from the non-commutativity of the $\widehat{P}_I$.

The coexact sector of Hilbert space is labelled by the usual photon occupation numbers. In the harmonic sector the momenta do not commute, so cannot simultaneously be diagonalized to label states by integer electric fluxes. Instead, the Hilbert space organizes itself into $g$ decoupled simple harmonic oscillators of frequency $\frac{|k|e^2}{2\pi}$. The degeneracy of each fixed set of oscillator occupation numbers is $|k|^g$ due to the compactness of $Q^I$. Altogether, we get the Hilbert space
\be
    \cH_{\text{MCS}} = \mathcal{F}_{\text{co}}\otimes L^2 \left(H^1(\Sigma_2,U(1)),\mathscr{L}_k \right) \cong \mathcal{F}_{\text{co}} \otimes \mathcal{F}_{\text{h}} \otimes \mathcal{H}_{U(1)_k \text{CS}} \, .
\ee
$\cF_{\text{co}}$ is the (massive) photon Fock space. The harmonic sector Hilbert space decomposes into $\cF_{\text{h}}$, the Fock space of the $g$ simple harmonic oscillators, and $\mathcal{H}_{U(1)_k \text{CS}}$ that captures the degeneracy of each fixed set of oscillator occupation numbers, and coincides with the Hilbert space of pure Chern-Simons theory. A useful labeling of basis states in the Hilbert space is given by
\be
    | \{ N_r \}_{r \in \bN} ; \{ n_i \}_{i=1,\cdots ,g} ; \{ l_i \}_{i=1,\cdots ,g} \rangle \, , \qquad N_r,n_i \in \bZ_{\ge 0} \, , \quad l_i \in \bZ_{|k|} \, .
\ee
The $k^{2g}$-many operators
\be \label{eq: 1fs ops in MaxCS}
    U_{\bold{n}} := \exp \left(\frac{2\pi i}{k} n_I (\cI^{-1})^{JI} \widehat{P}_J + in_I \widehat{Q}^I \right) \, , \qquad \bold{n} \in \bZ_{|k|}^{2g} \, ,
\ee
are the generators of the $\bZ_{|k|}$ 1-form symmetry. One can verify that $[\mathopen{:}\widehat{H}_{\text{MCS}}\mathclose{:},U_{\bold{n}}]=[\widehat{P}_I,U_{\bold{n}}]=0$. Since they commute with $\widehat{P}_I$, they further preserve the harmonic sector eigenvalues $\{ n_i \}$, meaning they act within the $|k|^g$-dimensional degeneracy states. These operators comprise the $|k|^{2g}$ dimensional algebra of operators of the $U(1)_k$ Chern-Simons theory. 

\paragraph{BF-Chern-Simons theory.} Consider now BF-Chern-Simons theory \cite{Witten:1988hf,Elitzur:1989nr,Horowitz:1989ng,Birmingham:1991ty}:
\be
    S_{\text{BFCS}} = -\frac{1}{2\pi} \int_{M_3} b_1 \wedge \d A_1 - \frac{k}{4\pi} \int_{M_3} A_1 \wedge \d A_1 \, ,
\ee
the Lagrange multiplier constraints give $\d A=\d b=0$ after a suitable linear combination. The mode algebra is again \eqref{eq: Harmonic mode algebra with K}. The phase space is the cotangent bundle of the holonomy torus $T^*U(1)^{b_{1}(\Sigma_2)}$, with modified symplectic form that receives a contribution from the Chern-Simons term. Geometric quantization in the vertical polarization, along the cotangent fibers, gives wavefunctions that depend on the holonomies and are square-integrable sections of the prequantum line bundle $\mathscr L_k$ over the holonomy torus \cite{Axelrod:1989xt}:
\be\label{eq: BFCS Hilbert space}
\cH_{\mathrm{BFCS}}=
L^2\!\left(H^1(\Sigma_2,U(1)),\mathscr L_k\right).
\ee
A convenient choice of basis for this Hilbert space is to label states by $| \{ n_i \}_{i=1}^g ; \{ l_i \}_{i=1}^g \rangle$ with $n_i \in \bZ_{\ge 0}$ and $l_i \in \bZ_{|k|}$, as in Maxwell-Chern-Simons theory. The algebra of operators on this Hilbert space is generated by
\be
W_{\bold n}=e^{in_I\widehat Q^I} \, ,\qquad
V_{\boldsymbol\alpha}=e^{2\pi i\alpha^I\widehat P_I} \, ,
\qquad\bold n\in\bZ^{2g},\quad\boldsymbol\alpha\in\bR^{2g} \, .
\ee
To state the label identifications, consider the Weyl-ordered operators
\be\label{eq: BFCS canonical lines}
L_{\bold n,\boldsymbol\alpha}
=\exp\left(in_I\widehat Q^I+2\pi i\alpha^I\widehat P_I\right) \, .
\ee
For $\boldsymbol\ell\in\bZ^{2g}$, we use the convention $L_{k\cI\boldsymbol\ell,\boldsymbol\ell}=\mathbf 1$, where $(\cI\boldsymbol\ell)_I=\cI_{IJ}\ell^J$. The operator labels are therefore identified, up to an overall phase, as
\be\label{eq: BFCS line identification}
(\bold n,\boldsymbol\alpha)
\sim(\bold n+k\cI\boldsymbol\ell,\boldsymbol\alpha+\boldsymbol\ell) \, ,
\qquad\boldsymbol\ell\in\bZ^{2g} \, .
\ee
With the Weyl-ordering convention of \eqref{eq: BFCS canonical lines}, the exact operator relation is\footnote{The phase follows from \eqref{eq: Harmonic mode algebra with K} by the Baker-Campbell-Hausdorff formula. A spacetime line $e^{i\oint_{\gamma^I}(nA_1+\alpha b_1)}$ on a fixed cycle $\gamma^I$ has canonical labels $n_J=n\delta_{IJ}$ and $\alpha^J=-\alpha(\cI^{-1})^{IJ}$. Thus \eqref{eq: BFCS line identification} reduces to the spacetime identification $(n,\alpha)\sim(n+k,\alpha+1)$ with no extra phase.}
\be\label{eq: BFCS Weyl phase}
L_{\bold n+k\cI\boldsymbol\ell,\boldsymbol\alpha+\boldsymbol\ell}
=e^{i\pi n_I\ell^I}L_{\bold n,\boldsymbol\alpha} \, .
\ee
These operators $\{ U_{\bold{n}},V_{\boldsymbol{\alpha}} \}$ generate the $(\bZ_{|k|}\times \bR)$ 1-form symmetry group of BF-Chern-Simons theory. 

\subsection{Comments on Canonical Quantization with Boundary}

Let us briefly comment on what happens to the discussion of the previous subsections in the presence of a boundary, that is relevant for both the holographic and the SymTFT framework. Canonical quantization on a manifold with boundary works in much the same way as on a closed manifold, with the only additional data being edge modes and electric or magnetic fluxes dressing the boundary. For the purposes of our later discussion, we only need to understand the homogeneous Dirichlet and Neumann boundary conditions, chosen to ensure a well-defined variational principle. See \cite{Kapustin:2009av,DiPietro:2019hqe,Arbalestrier:2025jsg} for detailed discussions on general Maxwell boundary conditions. 

Importantly for the use of the Hodge decomposition, one further requires additional boundary conditions on top of the usual Dirichlet and Neumann conditions to ensure a self-adjoint Laplace operator. These are called relative and absolute boundary conditions, respectively. Note they are not additional physical restrictions, since the former is guaranteed by the Coulomb gauge, and the latter is an additional gauge fixing choice. Choosing either boundary conditions ensures that the canonically quantized theory does not possess boundary fluxes, and that dualizing Maxwell in the presence of the boundary does not acquire additional boundary terms \cite{Gaiotto:2008ak}. 

In the case of the relative boundary, constant boundary preserving gauge parameters parameterize the boundary $U(1)$ global symmetry. In the absolute case boundary gauge transformations are allowed and generate the $U(1)$ gauge symmetry of the boundary theory. In terms of the bulk, the presence of the boundary modifies the bulk fields to be decomposed according to the Hodge-Morrey-Friedrichs decomposition \cite{Schwarz1995} naturally adapted to the corresponding boundary condition. After performing this decomposition, the qualitative features of the quantization are analogous to the closed manifold case, with the only technical difference being the boundary conditions on the form spaces the exact, coexact, and harmonic sectors are decomposed onto. Thus we can safely apply the qualitative results of our earlier closed manifold canonical quantization in what follows.

\section{Continuous SymTFTs from Dynamical Theories} \label{sec: Cont symtft from dynamical theories}
In this Section we rigorously establish a relation, at the quantum level, between the SymTFT for a continuous symmetry, and the dynamical holographically dual gauge theory. For the sake of concreteness and simplicity we discuss this for a single non-anomalous $U(1)$ 0-form symmetry however, the conclusions are valid for any continuous Abelian symmetry. Non-Abelian symmetries are beyond the scope of our work.

\subsection{Continuous SymTFT as a Subsector of Maxwell Theory}

\label{sec: topological subsector}

Consider Maxwell theory in its first order formulation
\be \label{eq: Max first order action}
    S_{\text{Max}} =-\frac{1}{2\pi} \int_{M_{d+1}} b_{d-1} \wedge \d A_1+ \frac{e^2}{8 \pi^2} \int_{M_{d+1}} b_{d-1} \wedge \star_{d+1} b_{d-1}  \, ,
\ee
where $b_{d-1} \in \Omega^{d-1}(M_{d+1},\bR)$ is a globally defined form. We emphasize that this is \emph{not} a gauge field: no redundancy of $b_{d-1}$ leaves the action invariant. Integrating $b_{d-1}$ out one recovers the usual Maxwell action \eqref{eq: Maxwell action}. Classically, the $e \rightarrow 0$ limit\footnote{By the $e\rightarrow 0,\infty$ limits, we mean the limit taken with respect to a fixed characteristic length scale $L$ set by the scale of the geometry $M_{d+1}$.} yields the topological BF-theory. Quantum mechanically this is \textbf{not} true. In fact the quantization of BF-theory uses that $b_{d-1}$ is a (non-compact) gauge field, and this cannot be reached smoothly in the limit $e\rightarrow 0$. Nevertheless a relation between the theories exists, that can be stated and derived rigorously using the canonically quantized picture.

Recall the normal ordered Hamiltonian of Maxwell theory
\be \label{eq:Hmax}
\mathopen{:}\widehat{H}_{\text{Max}}\mathclose{:}
    = \sum_{n \in \bN} \sqrt{\mu_n} \widehat{a}^{\dagger}_n \widehat{a}_n + \frac{e^2}{2} \sum_{I,J=1}^{b_1(\Sigma_d)} L^{IJ} \widehat{P}_I \widehat{P}_J  + \frac{2\pi^2}{e^2} \sum_{I,J=1}^{b_2(\Sigma_d)} G_{IJ} \widehat{m}^I \widehat{m}^J \, .
\ee
In the limit $e \rightarrow 0$ we observe that:
\begin{enumerate}
    \item States with non-zero magnetic flux become infinitely energetic, meaning finite energy states $|\psi \rangle$ must belong to the zero flux sector $\widehat{m}^I |\psi \rangle=0$. 
    \item The energy of Wilson line states generated by $\widehat{Q}_I$ go to zero, meaning they become topological in this limit. These interpolate to the topological Wilson lines of BF theory. 
    \item Since the energy of the photon modes is independent of $e$, they remain in the spectrum.
\end{enumerate}
Hence the Maxwell Hilbert space in the $e\rightarrow 0$ limit contains the Hilbert space of BF theory, alongside an infinite tower of excited photon states that have non-zero energy $\sqrt{\mu _n}$ that depends on the geometrical feature of space: the $e \rightarrow 0$ limit of Maxwell theory is not topological. While this last statement is in some sense a tautology, it is, in some other sense, also subtle as at the level of the classical action the dependence on the metric disappears: quantization does not commute with the limit $e\rightarrow 0$. We can summarize this discussion with following expression
\be
    U(1)\text{-Maxwell}_{e} \xrightarrow{e\rightarrow 0 }  \bR \text{-Maxwell} \times U(1)\text{-BF} \, .
\ee
 That is, the zero coupling limit of the compact Maxwell theory gives the decoupled sectors of two theories: non-compact $\bR$-Maxwell (containing the infinite tower of photons), and the $U(1)$ BF theory (containing the Wilson lines and electric Gukov Witten operators). Non-compact Maxwell does not depend on a coupling $e$, since the absence of a Dirac quantization condition means the coupling can be absorbed via a field redefinition. 

Another way to see that photons are not projected out of Maxwell theory in the zero coupling limit is the existence of a gauge invariant observable that can detect them as $e \rightarrow 0$, namely:
\be \label{eq: R-Maxwell field strength}
    \star_{d+1} f_2 = \frac{1}{e^2}\star_{d+1} F_2 \, .
\ee
$F_2$ is the field strength of the $U(1)$-Maxwell theory, and as $e \rightarrow 0$, $\star_{d+1}F_2$ does indeed vanish, but the ratio in \eqref{eq: R-Maxwell field strength} does not. It becomes the field strength of the non-compact $\bR$-Maxwell theory that decouples from the Wilson lines in the zero coupling limit. Non-zero correlation functions of this operator detects the photon states.

There is a very well defined operation that we can apply to the $e\rightarrow 0$ quantum theory to produce a TQFT, which is projecting out the photon sector to isolate the topological BF theory. We emphasize that the operation we are going to describe makes sense as an operation on a quantum system at any finite $e$, but it does not produce a \emph{local} quantum theory unless it is performed at $e\rightarrow 0$ (or $e\rightarrow \infty$ as we will see below). We first define \textit{topological states} $| \psi \rangle$ to be those annihilated by the photon number operator
\be
    N |\psi \rangle = 0 \, , \qquad  N = \sum_{n \in \bN} a_n^{\dagger} a_n \, ,
\ee
and the \textit{topological operators} $\cO_i$ that satisfy\footnote{The last condition has the scope of fixing possible fake dependence on the metric arising from redefinition of some operator by a function explicitly depending on the metric.}
\be
    [N,\cO_i]|\psi \rangle = 0 \, , \qquad \langle \psi |[N,\cO_i] = 0 \, , \qquad \delta_g \cO_i = 0 \, ,
\ee
on topological states $|\psi \rangle$. These operators preserve the space of topological states $\cH_{\text{top}}$, and form an algebra $\cA_{\text{top}}$ that is closed under adjoints.

We claim that $\cH_{\text{top}}$ is the Hilbert space of a local and topological theory, that has $\cA_{\text{top}}$ as its algebra of operators. To prove this we need to show that the stress-tensor has vanishing expectation values on topological states $\langle \psi_1 | \mathopen{:}T_{\mu \nu} \mathclose{:} | \psi_2 \rangle=0$. This is obviously true for the $\mathopen{:}T_{00} \mathclose{:}$ component, that is the Hamiltonian \eqref{eq:Hmax}. For the other components one uses the expression in terms of electric and magnetic fields, and inserts the mode expansions including the harmonic sectors. As an example, for $T_{0i}$ one finds that on topological states
\be
    \mathopen{:}T_{0i} \mathclose{:} |\psi \rangle = \frac{i(-1)^{d-1}}{2}\sum_{m,n \in \bN} \left(\frac{\mu_m}{\mu_n}\right)^{1/4} h^{jk} (\star \d \varphi_n)_j (\star \varphi_m)_{ik} a_n^{\dagger} a_m^{\dagger} |\psi \rangle \, .
\ee
the important fact is that all terms are proportional to $a_n^{\dagger} a_m^{\dagger} |\psi \rangle$, and that is true also for the other components $T_{ij}$. This implies $\langle \psi_1 | \mathopen{:}T_{0i} \mathclose{:} | \psi_2 \rangle=0$. In turn, given that the $\cO_i$ preserve the space of topological states, correlation functions are metric independent:
\be
\ba
    \delta_g \langle \cO_1 \cdots  \cO_n \rangle = &\left\langle -\frac{1}{2} \int_{M_{d+1}} \d^{d+1}x  \sqrt{-g} \, \delta g^{\mu \nu} T_{\mu \nu} \cO_1  \cdots  \cO_n \right\rangle = 0 \, ,
\ea
\ee
The result is therefore a TQFT. In our case, one identifies the $\cO_i$ operators as the topological Wilson lines and electric Gukov Witten operators that comprise the topological BF theory\footnote{The correlation function between two Wilson lines on the worldlines $\gamma_1=\{(t,\bold{0}) \mid t\in [0,\tau]\}$ and $\gamma_2=\{(t,\bold{r}) \mid t \in [0,\tau]\}$, in $(d+1)\ge4$ dimensional Euclidean Maxwell, for large $\tau$ is 
\be
    \langle W_n(\gamma_1) W_{m}(\gamma_2) \rangle \sim \exp\left( -nm \frac{e^2 \Gamma(\frac{d-2}{2})\tau}{4 \pi^{d/2}r^{d-2}} -\frac{(n^2+m^2)e^2 \Gamma(\frac{d-1}{2})\tau}{4 \pi^{(d+1)/2}(d-2)\epsilon^{d-2} }\right) \, .
\ee
 The first term is the Coulomb interaction potential, and the second is the leading term of the self-energy in the regularization scheme of \cite{Srednicki:2007qs}. The interaction potential vanishes in the $e\rightarrow 0$ limit indicating that the correlation function  agrees with that calculated in the BF-theory. Moreover, the self-energies generated by the photons is removed in the fixed $\epsilon$, $e \rightarrow 0$ limit, consistent with the observation that the photons decouple in this limit.}.

This is analogous to the argument that topologically twisted supersymmetric theories are topological \cite{Witten:1988ze}. In such theories, the result of the topological twist is that the energy momentum tensor becomes $\cQ$-exact $T_{\mu \nu} = \{ \cQ, G_{\mu\nu} \}$, with $\cQ$ the twisted supercharge. Physical states are those annihilated by the supercharge, and physical operators $\cO$ in the twisted theory are $\cQ$-closed and metric independent. The variation with respect to the metric of any correlation function of physical operators therefore vanishes.

The entire analysis above is analogous for $p$-form Maxwell theory, meaning that the continuous SymTFT for a $U(1)^{(p)}$ symmetry arises from a similar zero coupling limit and topological truncation. In $(2+1)$d this analysis is also valid for non-zero Chern-Simons terms $k\ne 0$: the SymTFT for anomalous $U(1)$ 0-form symmetry arises as the topological subsector of the zero coupling limit $e \rightarrow 0$ of Maxwell-Chern-Simons.

\noindent \textbf{The infinite coupling limit.} A crucial observation from \eqref{eq:Hmax} is that the opposite limit $e\rightarrow \infty$ also behaves in an analogous manner: the local photon sector is untouched, all electrically charged states become infinitely heavy while the magnetic states become degenerate at zero energy generating a topological sub-sector. The same photon decoupling procedure described above produces a topological theory with states created by $\widehat{m}^I$. This Hilbert space is equivalent with the one of theory obtained by quantizing the BF action
\begin{equation}\label{eq:dual BF}
    S=\frac{i}{2\pi}\int \widetilde{b}_2 \wedge \d\widetilde{A}_{d-2} \ ,
\end{equation}
where, $\widetilde{b}_2$ is non-compact while $\widetilde{A}_{d-2}$ is compact.

In fact this limit can be immediately understood in the electric-magnetic dual formulation of Maxwell theory, that involves a $(d-2)$-form $U(1)$ gauge field  $\widetilde{A}_{d-2}$ and coupling
\begin{equation}
    \widetilde{e}^2=\frac{4\pi ^2}{e^2} \ .
\end{equation}
The $e\rightarrow \infty$ limit corresponds to $\widetilde{e}\rightarrow 0$, hence our analysis implies that after the photon decoupling this gives the BF theory \eqref{eq:dual BF}.

The existence of two different topological corners raises the important question: how do we know, from first principles, which is the correct one to obtain the SymTFT? Answering this question is of central importance and clarifies the precise relation between the continuous SymTFT and holography. As anticipated in the introduction, our main result is that the two topological corners are both correct: they are two SymTFTs for two boundary theories related by dynamical gauging, explaining the conjectured map of TQFTs in \cite{Antinucci:2024zjp}. To derive this answer, we need to further discuss these two limits, and in particular how the boundary conditions and the various defects behave under them. So for the moment we will pause and analyze both limits, returning this question  in Section \ref{sec: derivation from holography}.

\subsection{Map of Boundaries and Defects}

The two topological corners of Maxwell theory retain different boundary symmetries. To identify the SymTFT of a given boundary theory, we must therefore follow both the boundary condition and its symmetry through the limit. For 1-form Maxwell theory in $(d+1)$-dimensions, the results are summarized in Table \ref{tab: boundary limits}. Throughout, we will refer to a topological limit as one that includes both the coupling limit and photon projection of Section \ref{sec: topological subsector}.

\begin{table}[!ht]
\centering
\begin{tabular}{@{}lccc@{}}
\toprule
Maxwell boundary & Finite $e$ & $e\rightarrow 0$ & $e\rightarrow\infty$ \\
\midrule
Dirichlet & $U(1)^{(0)}$ & $U(1)^{(0)}$ & $\bZ^{(1)}$ \\
Neumann & $U(1)^{(d-3)}$ & $\bZ^{(d-2)}$ & $U(1)^{(d-3)}$ \\
\bottomrule
\end{tabular}
\caption{Boundary symmetries of Maxwell theory and of its two topological subsectors. The two limiting columns refer to the topological limit.}
\label{tab: boundary limits}
\end{table}

The Dirichlet boundary retains its $U(1)^{(0)}$ symmetry in the weak-coupling corner, whereas the Neumann boundary retains its magnetic $U(1)^{(d-3)}$ symmetry in the strong-coupling corner. These are the two descriptions relevant to dynamical gauging. The other corners instead acquire discrete symmetries generated by operators that were non-topological at finite coupling. We now derive this dictionary; Section \ref{sec: gauging from holography} uses it to formulate the dynamical gauging map.

\paragraph{Operators and boundary conditions at finite coupling.}
We assume $d>1$. First write the operators on a spatial slice, with $\star$ the Hodge operator there. The integration supports below are closed spatial cycles. Their spacetime continuations will be used when discussing boundaries and endpoints.
The electric and magnetic operators are
\be\label{eq: Maxwell electric operators}
    W_n = \exp\left(in\oint_{\Sigma_1} A\right) \, , \qquad
    U_\alpha = \exp\left(\frac{2\pi i\alpha}{e^2}\oint_{\Sigma_{d-1}}\star E\right) \, ,
    \qquad n\in\bZ \, , \quad\alpha\in\bR/\bZ \, ,
\ee
and
\be\label{eq: Maxwell magnetic operators}
    \widetilde W_m = \exp\left(im\oint_{\Sigma_{d-2}}\widetilde A_{d-2}\right) \, , \qquad
    \widetilde U_\beta = \exp\left(i\beta\oint_{\Sigma_2} B\right) \, ,
    \qquad m\in\bZ \, , \quad\beta\in\bR/\bZ \, .
\ee
Here $\widetilde A_{d-2}$ is the compact gauge field dual to $A_1$. At finite coupling, $U_\alpha$ and $\widetilde U_\beta$ are topological Gukov-Witten operators generating the bulk $U(1)^{(1)}$ and $U(1)^{(d-2)}$ symmetries. Wilson and 't Hooft operators $W_n$ and $\widetilde W_m$ are generically non-topological.

With Dirichlet boundary conditions, Wilson lines can end on charged boundary point operators, and $U_\alpha$ restricts to their $U(1)^{(0)}$ symmetry defect. Magnetic Gukov-Witten surfaces can also end, while 't Hooft operators can lie on the boundary. The latter remain non-topological at finite coupling and hence do not yet generate a $\bZ^{(1)}$ symmetry. In the diagrams, blue denotes generically non-topological operators and red denotes topological Gukov-Witten operators:

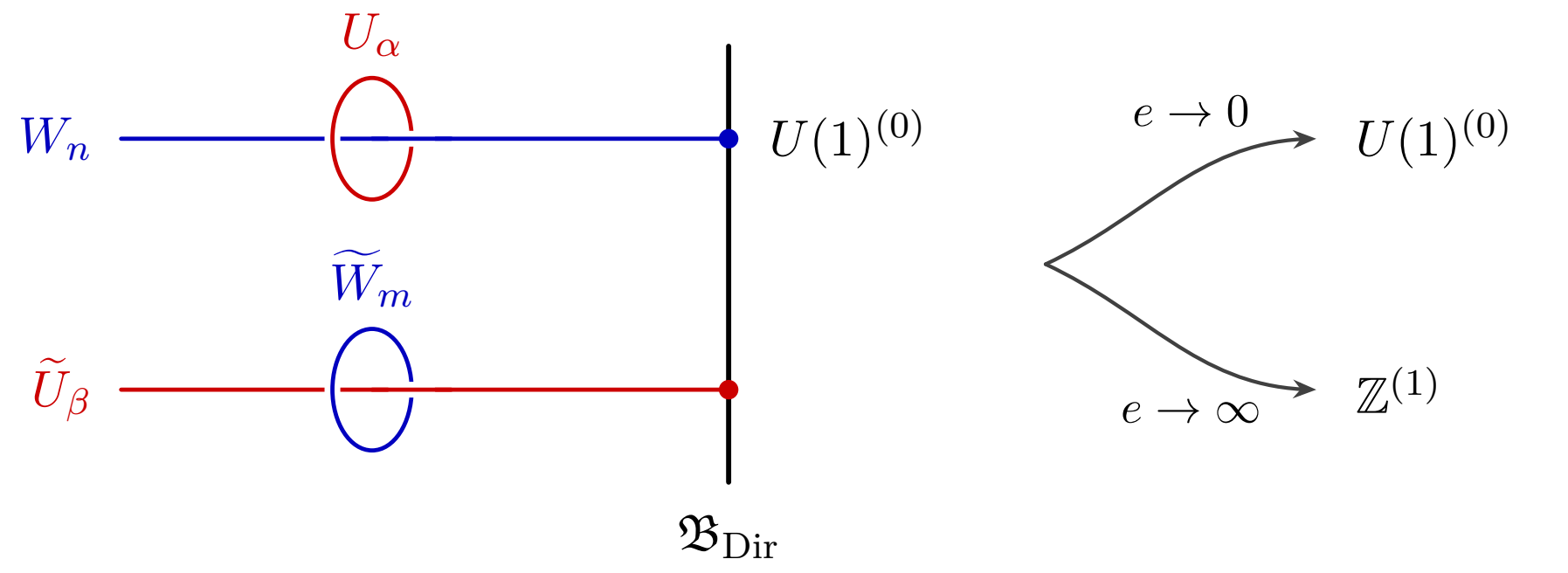
\begin{equation}
    \begin{tikzpicture}[
        x=1cm, y=1cm, baseline={(0,1.65)},
        line cap=round, line join=round,
        operator/.style={line width=0.9pt},
        crossing/.style={preaction={draw=white,line width=3.4pt}},
        limit/.style={draw=black!75,line width=0.8pt,
            -{Stealth[length=5pt,width=4pt]}},
        every node/.style={inner sep=2pt,font=\normalsize}
    ]
        \path[use as bounding box] (-1.0,-0.45) rectangle (11.25,3.65);
        \draw[black,line width=1.1pt] (4.6,0.10) -- (4.6,3.40);
        \node[anchor=north] at (4.6,-0.07) {$\mathfrak{B}_{\text{Dir}}$};

        \draw[operator,blue!75!black] (0,2.70) -- (4.6,2.70);
        \draw[operator,red!80!black,crossing]
            (1.9,2.70) ellipse [x radius=0.30,y radius=0.46];
        \draw[white,line width=3.4pt] (2.08,2.70) -- (2.32,2.70);
        \draw[operator,blue!75!black] (1.9,2.70) -- (2.5,2.70);
        \fill[blue!75!black] (4.6,2.70) circle [radius=2.1pt];
        \node[anchor=east,text=blue!75!black] at (-0.14,2.70) {$W_n$};
        \node[anchor=south,text=red!80!black] at (1.9,3.25) {$U_\alpha$};

        \draw[operator,red!80!black] (0,0.80) -- (4.6,0.80);
        \draw[operator,blue!75!black,crossing]
            (1.9,0.80) ellipse [x radius=0.30,y radius=0.46];
        \draw[white,line width=3.4pt] (2.08,0.80) -- (2.32,0.80);
        \draw[operator,red!80!black] (1.9,0.80) -- (2.5,0.80);
        \fill[red!80!black] (4.6,0.80) circle [radius=2.1pt];
        \node[anchor=east,text=red!80!black] at (-0.14,0.80) {$\widetilde U_\beta$};
        \node[anchor=south,text=blue!75!black] at (1.9,1.35) {$\widetilde W_m$};

        \node[anchor=west] at (4.83,2.70) {$U(1)^{(0)}$};
        \draw[limit] (7.0,1.75) to[out=25,in=180] (9.05,2.70);
        \draw[limit] (7.0,1.75) to[out=-25,in=180] (9.05,0.80);
        \node[font=\small,anchor=south] at (8.10,2.72) {$e\rightarrow0$};
        \node[font=\small,anchor=north] at (8.10,0.78) {$e\rightarrow\infty$};
        \node[anchor=west] at (9.27,2.70) {$U(1)^{(0)}$};
        \node[anchor=west] at (9.27,0.80) {$\bZ^{(1)}$};

    \end{tikzpicture} 
\end{equation}

Neumann boundary conditions exchange the electric and magnetic roles. The endpoints of 't Hooft operators carry the boundary $U(1)^{(d-3)}$ charge, measured by $\widetilde U_\beta$. Electric Gukov-Witten operators can end, while Wilson lines can lie on the boundary; these lines generate a $\bZ^{(d-2)}$ symmetry only when they become topological:

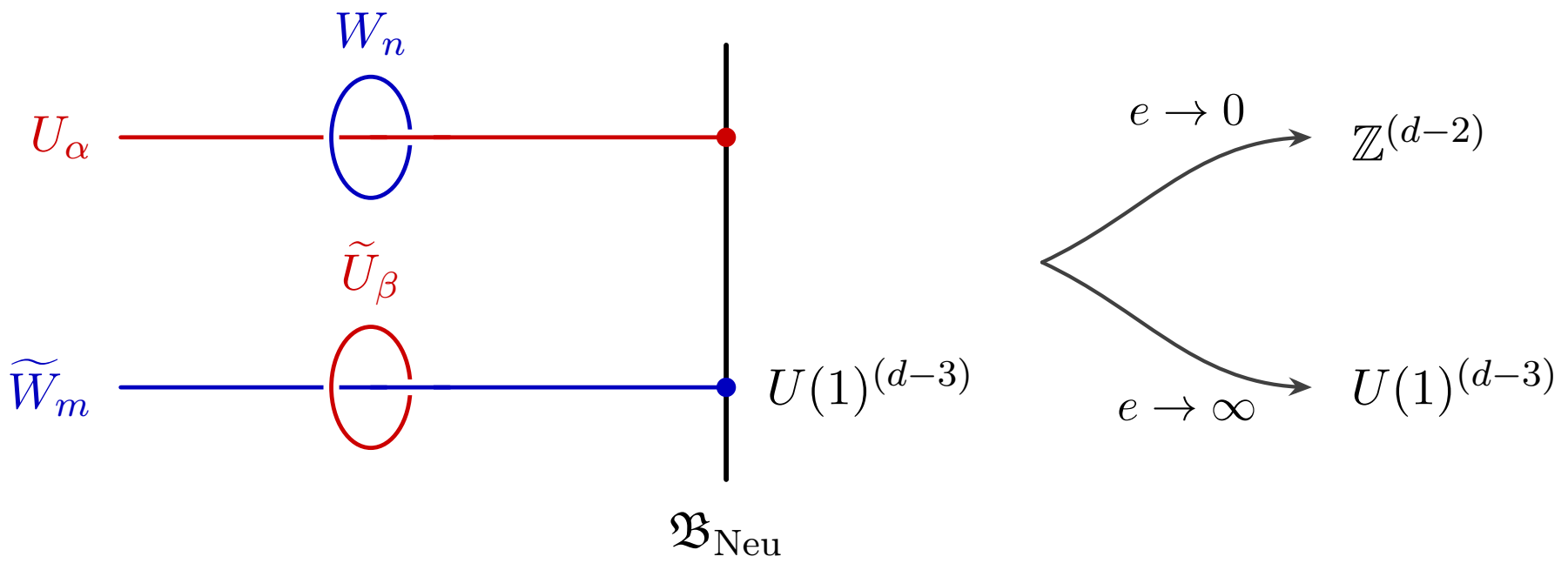
\begin{equation}
    \begin{tikzpicture}[
        x=1cm, y=1cm, baseline={(0,1.65)},
        line cap=round, line join=round,
        operator/.style={line width=0.9pt},
        crossing/.style={preaction={draw=white,line width=3.4pt}},
        limit/.style={draw=black!75,line width=0.8pt,
            -{Stealth[length=5pt,width=4pt]}},
        every node/.style={inner sep=2pt,font=\normalsize}
    ]
        \path[use as bounding box] (-1.0,-0.45) rectangle (11.25,3.65);
        \draw[black,line width=1.1pt] (4.6,0.10) -- (4.6,3.40);
        \node[anchor=north] at (4.6,-0.07) {$\mathfrak{B}_{\text{Neu}}$};

        \draw[operator,red!80!black] (0,2.70) -- (4.6,2.70);
        \draw[operator,blue!75!black,crossing]
            (1.9,2.70) ellipse [x radius=0.30,y radius=0.46];
        \draw[white,line width=3.4pt] (2.08,2.70) -- (2.32,2.70);
        \draw[operator,red!80!black] (1.9,2.70) -- (2.5,2.70);
        \fill[red!80!black] (4.6,2.70) circle [radius=2.1pt];
        \node[anchor=east,text=red!80!black] at (-0.14,2.70) {$U_\alpha$};
        \node[anchor=south,text=blue!75!black] at (1.9,3.25) {$W_n$};

        \draw[operator,blue!75!black] (0,0.80) -- (4.6,0.80);
        \draw[operator,red!80!black,crossing]
            (1.9,0.80) ellipse [x radius=0.30,y radius=0.46];
        \draw[white,line width=3.4pt] (2.08,0.80) -- (2.32,0.80);
        \draw[operator,blue!75!black] (1.9,0.80) -- (2.5,0.80);
        \fill[blue!75!black] (4.6,0.80) circle [radius=2.1pt];
        \node[anchor=east,text=blue!75!black] at (-0.14,0.80) {$\widetilde W_m$};
        \node[anchor=south,text=red!80!black] at (1.9,1.35) {$\widetilde U_\beta$};

        \node[anchor=west] at (4.83,0.80) {$U(1)^{(d-3)}$};
        \draw[limit] (7.0,1.75) to[out=25,in=180] (9.05,2.70);
        \draw[limit] (7.0,1.75) to[out=-25,in=180] (9.05,0.80);
        \node[font=\small,anchor=south] at (8.10,2.72) {$e\rightarrow0$};
        \node[font=\small,anchor=north] at (8.10,0.78) {$e\rightarrow\infty$};
        \node[anchor=west] at (9.27,2.70) {$\bZ^{(d-2)}$};
        \node[anchor=west] at (9.27,0.80) {$U(1)^{(d-3)}$};

    \end{tikzpicture}
\end{equation}

\paragraph{The weak-coupling corner.}
As $e\rightarrow0$, states with nonzero magnetic flux leave the finite-energy spectrum. After removing photon excitations, the surviving Wilson lines and electric Gukov-Witten operators are described by
\be\label{eq: continuous SymTFT}
    S=-\frac{1}{2\pi}\int_{M_{d+1}}b_{d-1}\wedge\d A_1 \, .
\ee
The magnetic symmetry acts trivially on this sector. For a Dirichlet Maxwell boundary, $A_1$ remains Dirichlet and the original $U(1)^{(0)}$ survives. For a Neumann Maxwell boundary, the now-topological Wilson lines generate $\bZ^{(d-2)}$, charging the endpoints of $U_\alpha$. The finite-coupling magnetic $U(1)^{(d-3)}$ is lost.

We can prove this last statement rigorously. On a closed spatial two-cycle, the period of $B$ equals that of its harmonic part $b_{\text{h}}$. Let $\cH_\Lambda$ be the spectral subspace with energy at most $\Lambda$, and take $e^2\Lambda\rightarrow0$ at fixed geometry. Positivity of the harmonic magnetic term in \eqref{eq:Hmax} gives, for normalized states in this subspace,
\be\label{eq: magnetic harmonic bound}
    \left\langle\int_{\Sigma_d}b_{\text{h}}\wedge\star b_{\text{h}}\right\rangle
    \leq 2e^2\left\langle\mathopen{:}\widehat H_{\text{Max}}\mathclose{:}\right\rangle
    \leq 2e^2\Lambda \, .
\ee
Pairing the flux with the harmonic representative of its Poincar\'e dual and applying Cauchy-Schwarz yields
\be
    \left|\oint_{\Sigma_2}B\right|^2
    =\left|\oint_{\Sigma_2}b_{\text{h}}\right|^2
    \leq C_{\Sigma_2}\int_{\Sigma_d}b_{\text{h}}\wedge\star b_{\text{h}} \, ,
    \qquad C_{\Sigma_2}=\|\eta_{\Sigma_2}\|_{L^2(\Sigma_d)}^2 \, .
\ee
Thus, using $2-2\cos x\leq x^2$,
\be\label{eq: magnetic symmetry decoupling}
\ba
    \|\bigl(\widetilde U_\beta(\Sigma_2)-1\bigr)|\psi\rangle\|^2
    &=\left\langle 2-2\cos\left(\beta\oint_{\Sigma_2}B\right)\right\rangle\cr
    &\leq \beta^2\left\langle\left(\oint_{\Sigma_2}B\right)^2\right\rangle
    \leq 2\beta^2 C_{\Sigma_2}e^2\Lambda\underset{e\rightarrow0}{\longrightarrow}0 \, .
\ea
\ee
Flux quantization excludes nonzero magnetic sectors from this energy window for sufficiently small $e$. In the extracted BF theory, $\widetilde U_\beta$ therefore acts as the identity.

\paragraph{The strong-coupling corner.}
Electric-magnetic duality gives the second corner directly: $\widetilde e^2=4\pi^2/e^2$, so $e\rightarrow\infty$ is weak coupling for $\widetilde A_{d-2}$. Nonzero electric flux becomes infinitely costly, $U_\alpha$ acts trivially on the retained sector, and photon decoupling leaves
\be\label{eq: dual SymTFT}
    \widetilde S=-\frac{1}{2\pi}\int_{M_{d+1}}\widetilde b_2\wedge\d\widetilde A_{d-2} \, .
\ee
Its operators are the surviving topological 't Hooft and magnetic Gukov-Witten operators. Maxwell Neumann becomes Dirichlet for $\widetilde A_{d-2}$, preserving $U(1)^{(d-3)}$. Maxwell Dirichlet instead becomes the Neumann boundary of this dual BF theory: the 't Hooft operators lying on the boundary become topological and generate $\bZ^{(1)}$.

The distinction in Table \ref{tab: boundary limits} is therefore essential. Changing Dirichlet to Neumann in the dynamical Maxwell theory implements dynamical gauging on the boundary, but extracting the symmetry before and after that operation selects different BF theories. Changing Dirichlet to Neumann within either BF theory instead gives the corresponding discrete symmetry, corresponding to flat gauging on the boundary. The next subsection explains the boundary meaning of the operators in this map.

\subsection{Perspective from the Boundary} \label{sec: boundary perspective}

To interpret Maxwell theory and its two topological corners as symmetry bulks, we first identify the relevant operators directly in a $d$-dimensional theory with a $U(1)^{(0)}$ symmetry and in its gauging. We then give their Dirichlet and Neumann Maxwell realizations at finite coupling, and follow them through the two limits. Here $X_d=\partial M_{d+1}$ is boundary spacetime and $\star$ is its Hodge star. We assume $d\geq3$ for the ordinary monopole endpoint description; at $d=2$ the dual symmetry is a $(-1)$-form symmetry and requires separate treatment.

\paragraph{The boundary theory and its gauging.}
Consider first a $d$-dimensional theory with a non-anomalous $U(1)^{(0)}$ symmetry, independently of any bulk description. Familiar objects include the charged point operators $\cO_n$ and the symmetry defects
\be
    U_\alpha(\Sigma_{d-1})=\exp\left(2\pi i\alpha\oint_{\Sigma_{d-1}}\star j\right),
    \qquad \d\star j=0,\qquad \alpha\in\bR/\bZ.
\ee
The other objects involve the background connection $\cA_1$ and its curvature $\cF_2$. On a contractible patch away from charged insertions, current conservation allows a local trivialization $\eta_{d-2}$ of $\star j$. The current coupling can then be written as
\be\label{eq: boundary curvature coupling}
    \star j=\d\eta_{d-2},\qquad
    S_{\text{mc}}=i\int_{X_d}\cA_1\wedge\star j
    \overset{\mathrm{loc.}}{=}i\int_{X_d}\cF_2\wedge\eta_{d-2} \ .
\ee
This makes the endpoint of a symmetry defect explicit. Wherever the trivialization is available, Stokes' theorem gives
\be\label{eq: boundaries of 0-form top ops}
    U_\alpha(\Sigma_{d-1})
    =\exp\left(2\pi i\alpha\int_{\Sigma_{d-1}}\d\eta_{d-2}\right)
    =\exp\left(2\pi i\alpha\int_{\partial\Sigma_{d-1}}\eta_{d-2}\right).
\ee
Here $\alpha$ is a chosen real lifts: integer shifts can change the flux prescription or attach a properly quantized holonomy. A \emph{monodromy defect} \cite{Bianchi:2021snj} on $\Sigma_{d-2}=\partial\Sigma_{d-1}$ supplies this endpoint. The symmetry wall glues charge-$q$ fields by $\Phi_q|_+=e^{2\pi i q\alpha}\Phi_q|_-$. Equivalently, it is described by the singular background
\be\label{eq: monodromy background}
    \cA_1=2\pi\alpha\,\delta^{(1)}(\Sigma_{d-1}),\qquad
    \cF_2=2\pi\alpha\,\delta^{(2)}(\Sigma_{d-2}) \ .
\ee
 Locally one can instead use $\cA_1=\alpha\d\theta$ in the transverse plane: the connection is flat off the defect and its holonomy on a charge-$q$ probe is $e^{2\pi i q\alpha}$. Substituting this profile into \eqref{eq: boundary curvature coupling} localizes the current coupling on the defect,
\be\label{eq: monodromy current coupling}
    S_{\text{mc}}=2\pi i\alpha\int_{X_d}\delta^{(2)}(\Sigma_{d-2})\wedge\eta_{d-2}
    =2\pi i\alpha\int_{\Sigma_{d-2}}\eta_{d-2}.
\ee
Thus the singular background couples to the same local data that appears at the end of the symmetry wall in \eqref{eq: boundaries of 0-form top ops}.
For nonintegral $\alpha$, the defect is non-genuine: it requires the symmetry-wall attachment. At $\alpha=m\in\bZ$ the wall is trivial and the integral flux defect is genuine. 

\begin{figure}[t]
    \centering
    \begin{tikzpicture}
        \draw[thick, black] (0,0) -- (0,3);
        \draw[thick, black] (4,-1) -- (4,2);
        \draw (0,0) .. controls (1,-1) and (3,-1) .. (4,-1);
        \draw (0,3) .. controls (1,2) and (3,2) .. (4,2);
        \draw[ultra thick, color=red] (4.05,-1) -- (4.05,2);
        \filldraw[black] (2,0.5) circle (0pt) node[]{\Large{$\Sigma_{d-1}$}};
        \filldraw[black] (4.2,0.5) circle (0pt) node[anchor=west]{\Large{\textcolor{red}{$\Sigma_{d-2}$}}};
    \end{tikzpicture}
    \caption{A monodromy defect on $\Sigma_{d-2}$ (red), represented as the endpoint of a symmetry wall on $\Sigma_{d-1}$. Equivalently, the background connection has a branch cut across the wall.}
    \label{fig: monodromy defect}
\end{figure}
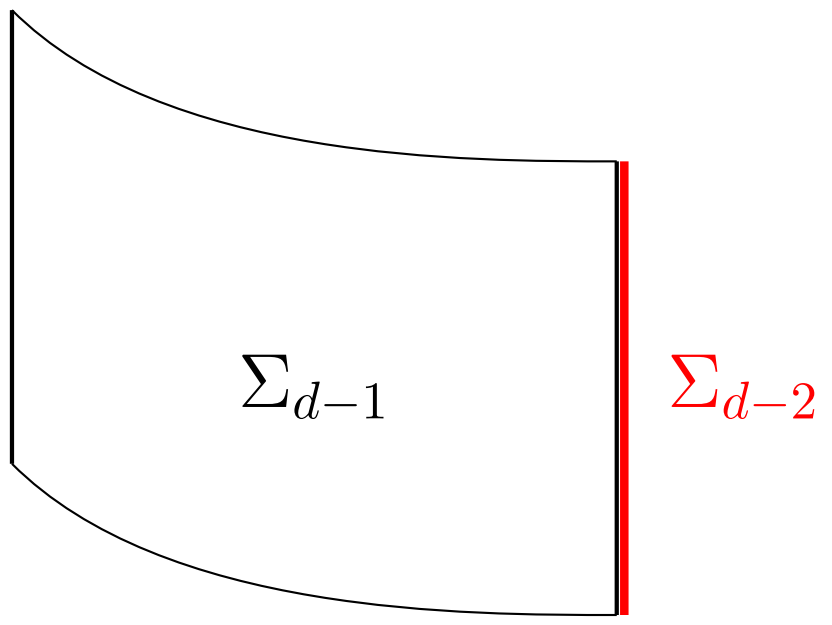

A \emph{background monopole} \cite{Kapustin:2011jm,Sachdev:2012tj,Pufu:2013eda} on $\gamma_{d-3}$ specifies a background with
\be\label{eq: Maxwell monopole operator}
    \oint_{S^2_{\mathrm{Link}}}\frac{\cF_2}{2\pi}=m\in\bZ,\qquad
    \d\cF_2=2\pi m\,\delta^{(3)}(\gamma_{d-3}).
\ee
This defines a \emph{family of insertions}, not a unique operator: the full curvature profile, and any necessary core data, must also be specified. It is generally a nonlocal background prescription, even when $\gamma_{d-3}$ is a point. For example, in spherical coordinates transverse to $\gamma_{d-3}$ one may choose the radial monopole background
\be\label{eq: radial background monopole}
    \cA_1=\frac{m}{2}(1-\cos\theta)\,\d\phi,\qquad
    \cF_2=\frac{m}{2}\sin\theta\,\d\theta\wedge\d\phi.
\ee
The potential is defined on a patch, while the curvature distributes flux $2\pi m$ over the linking sphere. Its gauge-dependent Dirac string is not a sheet of delta-function curvature.

\begin{figure}[t]
    \centering
    \begin{tikzpicture}
        \draw[thick, black] (0,0) -- (0,3);
        \draw[thick, black] (4,-1) -- (4,2);
        \draw (0,0) .. controls (1,-1) and (3,-1) .. (4,-1);
        \draw (0,3) .. controls (1,2) and (3,2) .. (4,2);
        \draw[ultra thick, color=red] (4.05,-1) -- (4.05,2);
        \filldraw[color=blue] (4.05,2) circle (3pt); 
        \filldraw[black] (2,0.5) circle (0pt) node[]{\Large{$\Sigma_{d-1}$}};
        \filldraw[black] (4.2,0.5) circle (0pt) node[anchor=west]{\Large{\textcolor{red}{$\Sigma_{d-2}$}}};
        \filldraw[black] (4.2,2) circle (0pt) node[anchor=west]{\Large{\textcolor{blue}{$\gamma_{d-3}$}}};
    \end{tikzpicture}
    \caption{The delta-curvature representative of a background monopole: $\gamma_{d-3}$ (blue) is the endpoint of the integral monodromy defect on $\Sigma_{d-2}$ (red). A general monopole profile with the same flux need not have this geometry.}
    \label{fig: background monopole}
\end{figure}
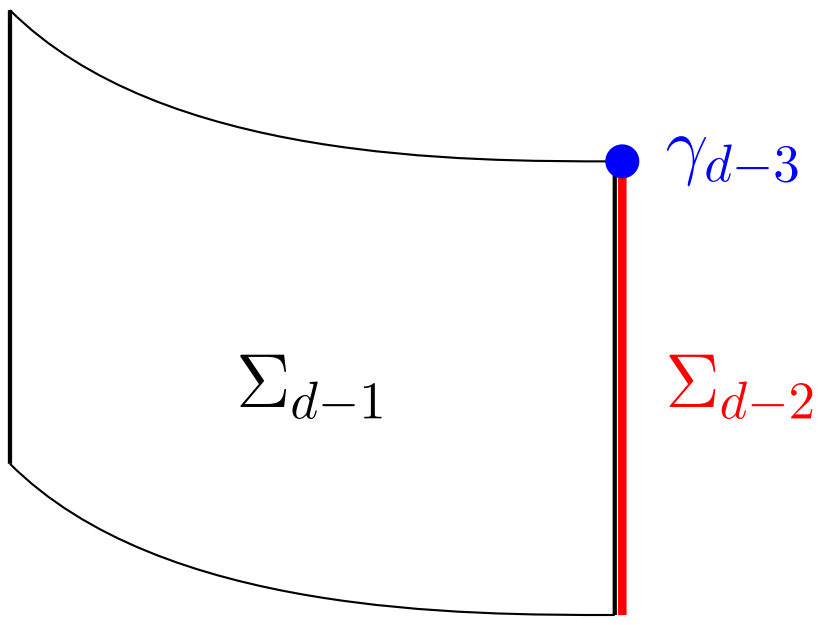

A different choice concentrates the same flux on an open integral monodromy defect ending on $\gamma_{d-3}$:
\be\label{eq: delta curvature monopole}
    \cF_2=2\pi m\,\delta^{(2)}(\Sigma_{d-2}).
\ee
We are taking $\partial \Sigma_{d-2}=\gamma_{d-3}$ so that $\d\delta^{(2)}(\Sigma_{d-2})=\delta^{(3)}(\gamma_{d-3})$. To check the monopole charge, take a small transverse ball $B^3_{\mathrm{Link}}$ meeting $\gamma_{d-3}$ once with positive intersection:
\be\label{eq: delta monopole flux}
    \oint_{S^2_{\mathrm{Link}}}\frac{\cF_2}{2\pi}
    =\int_{B^3_{\mathrm{Link}}}\frac{\d\cF_2}{2\pi}
    =m\int_{B^3_{\mathrm{Link}}}\delta^{(3)}(\gamma_{d-3})=m,
    \qquad S^2_{\mathrm{Link}}=\partial B^3_{\mathrm{Link}}.
\ee
The endpoint obeys the monopole condition \eqref{eq: Maxwell monopole operator}. Its local current coupling is \eqref{eq: monodromy current coupling} with $\alpha=m$ and an open support. This establishes the geometry in Figure \ref{fig: background monopole} for the delta-curvature representative. A general profile of the same charge need not admit that geometric description.

It is worth pointing out that, if the $U(1)^{(0)}$ symmetry happens to become trivially acting in some limit (e.g in the IR, or for some large value of a parameter), then the definition of the background monopole depends only on the class $\left[ \frac{\cF_2}{2\pi} \right]\in H^2(X_d,\bZ)$, and the family collapses to a single operator. To see this, observe that if the symmetry is trivially acting then $\star j=0$ and so $\d \eta_{d-2}=0$, meaning $\eta_{d-2}$ defines a class in cohomology $
    [\eta_{d-2}] \in H^{d-2}(X_d,\bR)$. Thus the monopole operator \eqref{eq: Maxwell monopole operator} depends only on the class 
$    [\Sigma_{d-2}] \in H_{d-2}(X_d,\bZ)$.By Poincar\'e duality, the integral therefore only depends on the cohomology class $    \left[ \frac{\cF_2}{2\pi} \right] \in H^2(X_d,\bZ) $.

Finally, a \emph{background Wilson line} is a c-number multiplying the identity:
\be
    \exp\left(in\oint_{\gamma_1}\cA_1\right),\qquad n \in \bZ \, .
\ee
For nonintegral charge $\beta$, a surface attachment is required: its definition is $\exp(i\beta\int_{\Sigma_2}\cF_2)$ with $\partial\Sigma_2=\gamma_1$. These source functionals become one at zero background.

Gauging promotes $\cA_1$ to a dynamical field $a_1$, so the background profiles themselves are now integrated over. Including a background $\widetilde{\cA}_{d-2}$ for the dual symmetry, the gauged partition function has the schematic form
\be\label{eq: boundary gauging with source}
    Z_{\mathrm{gauged}}[\widetilde{\cA}_{d-2}]
    =\int\cD a_1\,Z[a_1]e^{-S_{\mathrm{gauge}}[a_1]}
    \exp\left(\frac{i}{2\pi}\int_{X_d}\widetilde{\cA}_{d-2}\wedge f_2\right),
    \qquad f_2=\d a_1.
\ee
The integral includes the allowed bundles, and $S_{\mathrm{gauge}}$ supplies the gauge dynamics. The original closed symmetry walls act trivially on gauge-invariant states, while charged points require Wilson-line dressing. The Bianchi identity gives the dual $U(1)^{(d-3)}$ defects
\be
    \widetilde U_\beta(\Sigma_2)=\exp\left(i\beta\oint_{\Sigma_2}f_2\right),
    \qquad \beta\in\bR/\bZ.
\ee
A monopole insertion now fixes the linking flux of the \emph{dynamical} field, with the profiles away from the core integrated over. It is magnetically charged: a linking $\widetilde U_\beta$ measures the phase $e^{2\pi i\beta m}$. 

Background Wilson lines similarly become dynamical Wilson lines. They play the role of integral monodromy lines for the dual symmetry, as can be seen directly from \eqref{eq: boundary gauging with source}. Locally, integration by parts rewrites its background phase using $\widetilde{\cF}_{d-1}=\d\widetilde{\cA}_{d-2}$. A delta-curvature profile then gives
\be\label{eq: dual background Wilson attachment}
\ba
    \widetilde{\cF}_{d-1}&=2\pi n\,\delta^{(d-1)}(\gamma_1),\cr
    \exp\left(\frac{i}{2\pi}\int_{X_d}a_1\wedge\widetilde{\cF}_{d-1}\right)
    &=\exp\left(in\int_{\gamma_1}a_1\right)=W_n(\gamma_1).
\ea
\ee
If $\gamma_1$ is open, its endpoints carry opposite background-monopole charges for the dual symmetry, measured by $\oint_{S^{d-1}}\widetilde{\cF}_{d-1}/(2\pi)$. Gauge invariance requires charged insertions at these endpoints: for a line oriented from $x$ to $y$, the invariant combination is $\cO_n(x)W_n(\gamma_1)\cO_{-n}(y)$. This explains in what sense the original charged points play the role of dual-background monopoles. As before, fixed flux alone does not specify a general dual-background profile; the Wilson-line geometry selects its delta-curvature representative.

\paragraph{The Maxwell realization at finite coupling.}
For Dirichlet Maxwell, $A_1|_{X_d}=\cA_1$, a bulk $W_n$ ends on a charged point, whereas its restriction to the boundary is the background Wilson line. The restriction of $U_\alpha$ is the symmetry defect and its endpoint is the nonintegral monodromy defect. A bulk $\widetilde W_m$ lying on the boundary gives an integral monodromy defect: when it ends there, it fixes the background-monopole charge. The Dirichlet source must additionally supply the curvature profile. Only \eqref{eq: delta curvature monopole} makes this the endpoint of an integral monodromy defect lying wholly on the boundary. Finally, $\widetilde U_\beta$ ending on the boundary gives the attached background holonomy just described.

Neumann fixes the dual field $\widetilde A_{d-2}|_{X_d}=\widetilde{\cA}_{d-2}$ and makes $A_1|_{X_d}$ dynamical. The roles are exchanged: $\widetilde U_\beta$ gives the magnetic symmetry and $\widetilde W_m$ ends on its charged operators. $W_n$ lying on the boundary is the dynamical Wilson line, with dual-background monopoles as possible endpoints. Nonintegral monodromy lines of the dual symmetry are endpoints of $\widetilde U_\beta$. The remaining insertions are dual-background holonomies:
\be
    \widetilde W_m\big|_{X_d}=\exp\left(im\oint_{\Sigma_{d-2}}\widetilde{\cA}_{d-2}\right),\qquad
    \exp\left(i\alpha\int_{\Sigma_{d-1}}\widetilde{\cF}_{d-1}\right),
\ee
where the second expression describes an endpoint of $U_\alpha$, with its attachment explicit. The same bulk operators that describe Dirichlet monodromy defects therefore give source-only insertions on Neumann. This comparison does not retain an independently prescribed vortex singularity in the dynamical boundary field, and should not be confused with gauging that disorder prescription.

\paragraph{The two topological corners.}
The limits below include the finite-energy restriction and photon truncation of Section \ref{sec: topological subsector}. Table \ref{tab: boundary operators} separates the finite-coupling boundary dictionary from the surviving global operators.

At $e\rightarrow0$, $W_n$ becomes topological and $U_\alpha$ survives, while nonzero magnetic sectors are excluded and closed $\widetilde U_\beta$ act trivially. Dirichlet retains the original symmetry, its charged points and its symmetry-defects. Integral flux defects and monopoles do not furnish an additional magnetic sector of this SymTFT. On Neumann, the Wilson lines instead generate $\bZ^{(d-2)}$, charging the endpoints of $U_\alpha$, and the magnetic charged sector is lost.

At $e\rightarrow\infty$ the roles reverse: $\widetilde W_m$ becomes topological, $\widetilde U_\beta$ survives, and nonzero electric sectors are excluded. On Neumann this retains the magnetic symmetry and its charged operators. On Dirichlet, the integral monodromy defects generate $\bZ^{(1)}$ and charge the end-lines of $\widetilde U_\beta$. Their background-monopole endpoints have the topological attachment \eqref{eq: delta curvature monopole}. 
\begin{table}[!ht]
\centering
\small
\setlength{\tabcolsep}{4pt}
\begin{tabular}{@{}p{0.09\textwidth}>{\raggedright\arraybackslash}p{0.28\textwidth}>{\raggedright\arraybackslash}p{0.28\textwidth}>{\raggedright\arraybackslash}p{0.12\textwidth}>{\raggedright\arraybackslash}p{0.12\textwidth}@{}}
\toprule
Bulk & Dirichlet: restriction; endpoint & Neumann: restriction; endpoint & $e\to0$ & $e\to\infty$ \\
\midrule
$W_n$ & Background Wilson; charged point & Wilson line; dual background monopole & Topological & Excluded \\[4pt]
$U_\alpha$ & Symmetry defect; monodromy defect & Background flux phase; attached holonomy & Survives & Trivial action \\[4pt]
$\widetilde W_m$ & Integral monodromy; background monopole & Background holonomy; charged monopole & Excluded & Topological \\[4pt]
$\widetilde U_\beta$ & Background flux phase; attached holonomy & Magnetic symmetry; monodromy line & Trivial action & Survives \\
\bottomrule
\end{tabular}
\caption{Maxwell operators at the boundary and in the two topological corners. Each boundary condition entry lists the restriction and a possible endpoint of the bulk operators on the boundary. The last column refers exclusively to the behavior of the bulk operators $W_n,U_{\alpha},\widetilde{W}_m,\widetilde{U}_{\beta}$ in the limit, not to their restrictions or endpoints. }
\label{tab: boundary operators}
\end{table}

\section{The Dynamical Gauging Map}

\label{sec: gauging from holography}
In ordinary holography, dynamical gauging of a boundary continuous symmetry changes the boundary conditions of the bulk gauge field \cite{Witten:2003ya, Marolf:2006nd, DeWolfe:2020uzb}. Its action on symmetry data, however, relates distinct SymTFTs \cite{Antinucci:2024zjp}. We now unify these two pictures and we make the second precise, by using the coupling limits, photon decoupling, and boundary-operator dictionary of Section \ref{sec: Cont symtft from dynamical theories}. We first treat a single non-anomalous symmetry, then consider anomaly couplings. The latter refine the prescription: a topological limit of the bulk need not describe an admissible gauging of the boundary theory. The examples below explain how the local spectrum, the retained global sectors and the boundary anomaly together distinguish these possibilities.
\subsection{Derivation from Holography} \label{sec: derivation from holography}

Consider a non-anomalous $U(1)^{(0)}$ symmetry in $d\geq 2$ dimensions, with Maxwell theory on $M_{d+1}$ as its dynamical holographic description. With Dirichlet boundary conditions, $A_1|_{X_d}$ is the background field for the boundary symmetry. Integrating over this boundary value implements dynamical gauging, as in \eqref{eq: boundary gauging with source}, and replaces Dirichlet by Neumann boundary conditions \cite{Witten:2003ya,Marolf:2006nd}. A boundary gauge kinetic term gives a modified Neumann condition \cite{Apruzzi:2024htg}. Neumann for $A_1$ is Dirichlet for its compact dual $\widetilde A_{d-2}$. The latter couples to the magnetic $U(1)^{(d-3)}$ symmetry of the gauged theory\footnote{For $d=2$ this is a "$(-1)-$form symmetry", corresponding to a shift of the $\theta$ angle.}.

Starting with the Dirichlet boundary condition, we now ask how to obtain the SymTFT of the original theory from its dynamical holographic description. The SymTFT must preserve the topological symmetry data while forgetting the non-universal information carried by the current. The relation between conserved currents and monodromy defects gives a criterion for doing so: the existence of the current is equivalent to the existence of the corresponding monodromy defects\footnote{A gauge invariant conserved current $j$ produces the topological defect $U_\alpha(\Sigma_{d-1})=e^{i\alpha \int_{\Sigma_{d-1}} \star j}$ for $\Sigma_{d-1}$ closed, and the monodromy defect if $\Sigma_{d-2}=\partial \Sigma_{d-1}\neq \emptyset$. Viceversa, while the existence of $U_\alpha(\Sigma_{d-1})$ does not imply the existence of a current as the integrand does not need to be gauge invariant, intuitively, the existence of the monodromy defect tells that one can cut open $U_\alpha(\Sigma_{d-1})$ without spoiling gauge invariance on the boundary. This can be turned into a rigorous proof for CFTs \cite{Gabai2026}.}. We should therefore look for a limit that makes the states created by these defects infinitely energetic, while retaining the closed topological defects that implement the symmetry.

The boundary dictionary of Section \ref{sec: boundary perspective} identifies these monodromy defects with bulk magnetic-flux insertions. Nonintegral monodromy is included: it creates states with nonintegral magnetic flux in the corresponding defect background. By \eqref{eq: Hamiltonian of Maxwell}, the magnetic-flux contribution to their energy grows as $e^{-2}$, as is explicit for integral flux sectors in \eqref{eq:Hmax}. Hence states carrying a fixed nonzero magnetic flux become infinitely energetic at $e\rightarrow0$, for integral and nonintegral monodromy alike. This selects the weak-coupling limit for the original Dirichlet boundary theory. At the same time, the electric-flux energies vanish: the Wilson lines become topological and still end on the charged boundary operators, while $U_\alpha$ continues to implement the original $U(1)^{(0)}$ symmetry. The coupling limit alone does not give a TQFT, since the photon frequencies remain $\sqrt{\mu_n}$. Restricting to the photon vacuum and to the operators preserving it, as in Section \ref{sec: topological subsector}, completes the construction of the SymTFT.

With Neumann boundary conditions, we instead want to retain the magnetic $U(1)^{(d-3)}$ symmetry and its charged operators. For $d>2$, the bulk magnetic defects $\widetilde W_m$ can end on charged boundary operators, whose charge is measured by $\widetilde U_\beta$. The Hamiltonian \eqref{eq:Hmax} shows that the magnetic-flux energies vanish at $e\rightarrow\infty$, while nonzero electric-flux states become infinitely energetic. In this limit, $\widetilde W_m$ becomes topological and $\widetilde U_\beta$ survives, giving the magnetic symmetry algebra that we want to retain.

The same choice follows from the dual description: Neumann for $A_1$ is Dirichlet for $\widetilde A_{d-2}$, whose coupling obeys $\widetilde e^{\,2}=4\pi^2/e^2$. Thus $e\rightarrow\infty$ is the dual weak-coupling limit. For $d>2$, monodromy defects for the magnetic symmetry create states carrying magnetic flux of the dual field, or electric flux of the original field. Their energy diverges in this limit, implementing the same criterion of forgetting current information. The photon frequencies remain unchanged, so photon decoupling is again required to obtain the SymTFT.

Thus Dirichlet with $e\rightarrow0$ gives the SymTFT of the original theory, while Neumann with $e\rightarrow\infty$ gives that of the dynamically gauged theory, in both cases after photon decoupling. The other two boundary/limit combinations in Table \ref{tab: boundary limits} instead describe the discrete symmetries associated with flat gauging. On a closed spatial manifold, the two selected limits retain the electric and magnetic global sectors, respectively, with Hilbert spaces
\be\label{eq: gauging retained Hilbert spaces}
\cH_{\mathrm{top}}^{e\to0}\cong L^2\!\left(H^1(\Sigma_d,U(1))\right),
\qquad
\cH_{\mathrm{top}}^{e\to\infty}\cong L^2\!\left(H^{d-2}(\Sigma_d,U(1))\right),
\ee
under the torsion-free assumptions of Section \ref{sec: CQ of Maxwell}. Together with the retained operators and boundary conditions, these identify the two BF theories. In Euclidean signature the gauging map is
\be\label{eq: holographic BF gauging map}
\left(\frac{i}{2\pi}\int b_{d-1}\wedge\d A_1,\ \Dir(A_1)\right)
\quad\longrightarrow\quad
\left(\frac{i}{2\pi}\int\widetilde b_2\wedge\d\widetilde A_{d-2},\ \Dir(\widetilde A_{d-2})\right).
\ee
The compact fields are $A_1$ and $\widetilde A_{d-2}$; their BF partners are $\bR$ gauge fields. This is precisely the $p=0$ specialization of \eqref{eq: intro gauging substitution}:
\be\label{eq: holographic gauging substitution}
b_{d-1}\longmapsto\d\widetilde A_{d-2},\qquad
\d A_1\longmapsto\widetilde b_2.
\ee
 The quantum analysis explains which sectors each BF theory retains. Thus the map relates two topological limits of one dynamical theory; it neither changes a boundary condition within a fixed BF theory nor identifies the two Hilbert spaces in \eqref{eq: gauging retained Hilbert spaces}. The same argument extends to non-anomalous $U(1)^{(p)}$ symmetries with $0<p\leq d-2$, using a bulk $(p+1)$-form Maxwell field and its $(d-p-2)$-form dual, and yields the general gauging map \eqref{eq: intro gauging substitution}.

The map was motivated through matter-photon SymTFT coupling and operator endability in \cite{Antinucci:2024zjp}, and a holographic realization was proposed in \cite{Mignosa:2025cpg}. Here its physical origin is made explicit by the separation of photon and global sectors. Near-boundary limits in AdS can induce related coupling scalings, with applicability depending on dimension, form degree and boundary conditions \cite{Bonetti:2024cjk,Bergman:2025isp,Mignosa:2025cpg}.

This is the general organizing principle: choose an admissible boundary gauging, identify the symmetry data to retain, and restrict to the corresponding topological sector. Additional couplings affect both admissibility and the limiting spectrum. The next examples supply the extra rules needed to implement the principle beyond the non-anomalous single-$U(1)$ case.

\subsection{Anomalous Symmetries}
\label{sec: anomalous gauging}

Anomalies constrain which bulk boundary conditions implement ordinary gauging. In the quadratic three-dimensional examples, the obstruction to gauging is also reflected in the limiting local spectrum. We call a coupling limit \textit{smooth} in the local sector when the photon frequencies at each fixed spatial mode remain finite before any topological restriction. This does not require the harmonic sector to remain unchanged: its reorganization is part of the gauging map. A limit in which the photons become infinitely heavy can still define a consistent bulk TQFT, but not necessarily a SymTFT: the latter also requires a topological symmetry boundary. The chiral Chern-Simons theory in the first example illustrates this obstruction. Such a limit differs from a smooth coupling limit followed by restriction to the topological sector. The cubic five-dimensional example will require a separate, action-level analysis. 

\subsubsection{A Pure Anomaly: Maxwell-Chern-Simons Theory}
\label{sec: pure anomaly example}

Our first example is Maxwell-Chern-Simons theory, that holographically describes a two-dimensional $U(1)$ symmetry with anomaly coefficient $k$. We take $k\in2\bZ\setminus\{0\}$ as in Section \ref{sec: MCS and BFCS CQ}. At the level of the classical action, we have the following interpolation of limits
\be\label{eq: MCS topological corners}
\begin{aligned}
e\rightarrow0:\quad&S_{\mathrm{BFCS}}=
\frac{i}{2\pi}\int_{M_3}b_1\wedge\d A_1+
\frac{ik}{4\pi}\int_{M_3}A_1\wedge\d A_1 \, , \\
e\rightarrow\infty:\quad&S_{\mathrm{CS}}=
\frac{ik}{4\pi}\int_{M_3}A_1\wedge\d A_1 \, .
\end{aligned}
\ee
Their quantum origins are different. From \eqref{eq: MCS Hamiltonian}, the nonzero-mode photon frequencies and the harmonic oscillator frequency are
\be\label{eq: MCS mass and frequencies}
\omega_n=\sqrt{\mu_n+m_{\mathrm{CS}}^2}\, ,\qquad
m_{\mathrm{CS}}=\frac{|k|e^2}{2\pi}\, .
\ee
At weak coupling, $\omega_n\rightarrow\sqrt{\mu_n}$: the propagating sector remains and must be removed separately, as we explained in section \ref{sec: topological subsector}. All states with fixed harmonic oscillator occupation numbers, however, approach zero energy. Retaining this full harmonic space gives the BF-Chern-Simons Hilbert space described in Section \ref{sec: MCS and BFCS CQ}, with its Wilson and electric Gukov-Witten operators. With the Dirichlet symmetry boundary, it is the SymTFT of the original anomalous $U(1)$.

At strong coupling, both the nonzero-mode photons and the harmonic oscillator excitations become infinitely energetic. After subtracting the vacuum energy, only the $|k|^g$-dimensional degenerate ground space on a genus-$g$ surface remains. The operators \eqref{eq: 1fs ops in MaxCS} act on it as the Wilson-lines algebra of $U(1)_k$ Chern-Simons theory. Thus the second line of \eqref{eq: MCS topological corners} is a valid bulk limit, obtained by dynamically removing the massive sector \cite{Dunne:1998qy}.

It does not implement ordinary gauging of the anomalous boundary $U(1)$. The Chern-Simons term has boundary gauge variation
\be\label{eq: MCS boundary anomaly}
\delta_\lambda S_{\mathrm{CS}}=
\frac{ik}{4\pi}\int_{X_2}\lambda\,\d A_1\, ,
\qquad A_1\longmapsto A_1+\d\lambda \, .
\ee
Promoting the boundary connection to an ordinary gauge field therefore requires anomaly cancellation. The corresponding gauged symmetry boundary is obstructed. In the Maxwell description this obstruction accompanies the failure of a smooth local strong-coupling limit: the single mass $m_{\mathrm{CS}}$ necessarily diverges, and there is no second coupling with which to balance it. The existence of the Chern-Simons endpoint alone is consequently insufficient to infer a boundary gauging.

\subsubsection{A Mixed Anomaly in Two Dimensions}
\label{sec: mixed anomaly 2d}

Consider instead two $U(1)$ 0-form symmetries with a mixed anomaly and no pure anomalies. The holographic bulk action is
\be\label{eq: MaxMaxCS}
S=\frac{1}{2e_1^2}\int_{M_3}\d A_1^1\wedge\star\d A_1^1
+\frac{1}{2e_2^2}\int_{M_3}\d A_1^2\wedge\star\d A_1^2
+\frac{ik}{2\pi}\int_{M_3}A_1^1\wedge\d A_1^2 \, ,
\ee
where the upper indices label the two factors and lower indices denote form degree. We choose Dirichlet for $A_1^1$ and Neumann for $A_1^2$, thereby gauging $U(1)_2$. In the chosen convention, gauge invariance under $U(1)_2$ is preserved exactly, and the mixed anomaly appears in the response to $U(1)_1$ backgrounds.

For local fluctuations, canonical normalization $A_1^\alpha=e_\alpha a_1^\alpha$ makes the Chern-Simons coefficient proportional to $ke_1e_2$. The two propagating polarizations have opposite helicities with equal mass and classical mode frequency
\be\label{eq: mixed photon mass}
m_{12}=\frac{|k|e_1e_2}{2\pi} \, ,\qquad
\omega_n=\sqrt{\mu_n+m_{12}^2} \, , \qquad n \in \bN \,.
\ee
Unlike the pure-anomaly example, the boundary gauging admits the double scaling limit
\be \label{eq: balanced mixed limit}
e_1 \rightarrow 0 \, ,\qquad e_2 \rightarrow \infty \, ,\qquad
e_1e_2=c \, , \qquad 0<c<\infty \, .
\ee
The entire local spectrum remains fixed along this path. Gauging both factors is obstructed by their mixed anomaly, since taking both couplings to infinity makes $m_{12}$ diverge. This distinguishes a mixed anomaly from a pure anomaly of either factor. After gauging one factor, the other continuous symmetry is modified, and we will derive the finite remnant below.

This criterion has a simple general form for quadratic Abelian Maxwell-Chern-Simons theories with $U(1)^r$ gauge group. Writing their Chern-Simons term as $\frac{i}{4\pi}\int K_{\alpha\beta}A_1^\alpha\wedge\d A_1^\beta$, with $K$ symmetric and integral and diagonal Maxwell kinetic terms, the photon masses are the absolute values of the eigenvalues of
\be\label{eq: CS mass matrix}
\mathsf M_{\alpha\beta}=\frac{e_\alpha K_{\alpha\beta}e_\beta}{2\pi} \, ,
\ee
and helicities determined by their signs. Let $\Gamma \subset U(1)^r$ denote the $U(1)$-factors to be gauged, whose couplings tend to infinity, while the remaining couplings tend to zero. Bounded photon masses requires $K|_{\Gamma \times \Gamma}=0$, since any nonzero entry in this block would force the norm of $\mathsf M_{\alpha\beta}$ to diverge. When this block vanishes, finite double scaling limits of all remaining mixed products keeps all photon masses bounded. Thus the existence of a smooth limit in the local sector is equivalent to vanishing of the anomaly restricted to the subgroup selected by the limit.

\paragraph{The topological action and its operators.}
Before going into the quantum analysis, let us discuss a simple semi-classical argument to get intuition on what we should expect in the limit $e_1\rightarrow 0, e_2 \rightarrow \infty$. This has several limitations, notably it does not capture the necessity of the double scaling limit, but it is still useful to quickly get to the right answer (that we prove below). Observe that we can rewrite the action \eqref{eq: MaxMaxCS} in the first order form
\be\label{eq: mixed first order action}
S=\frac{e_1^2}{8\pi^2}\int_{M_3} b_1^1\wedge\star b_1^1
+\frac{1}{2e_2^2}\int_{M_3}\widetilde b_2^2\wedge\star\widetilde b_2^2
+\frac{i}{2\pi}\int_{M_3}\left[b_1^1\wedge\d A_1^1
+\widetilde b_2^2\wedge(\d\widetilde A_0^2+kA_1^1)\right] \, .
\ee
Here $\widetilde A_0^2$ is a $2\pi$-periodic scalar. Integrating it out makes $\widetilde b_2^2$ the curvature of a compact connection, while integrating out $b_1^1$ recovers the first Maxwell term. In the limit  $e_1\rightarrow 0, e_2 \rightarrow \infty$ the two quadratic coefficients vanish. The resulting topological action, whose Hilbert space and operators we recover below by restriction to the topological sector, is
\be\label{eq: mixed topological action}
S_{\mathrm{top}}=\frac{i}{2\pi}\int_{M_3}
\left[b_1^1\wedge\d A_1^1+
\widetilde b_2^2\wedge(\d\widetilde A_0^2+kA_1^1)\right] \, .
\ee
This theory is equivalent to the untwisted $\bZ_{|k|}$ gauge theory, i.e. the SymTFT of a non-anomalous $\bZ_{|k|}$ symmetry on the boundary. To see this, we look at the equations of motion
\be\label{eq: mixed topological equations}
\d A_1^1=0,\qquad\d\widetilde A_0^2+kA_1^1=0 \, ,\qquad
\d\widetilde b_2^2=0 \, ,\qquad\d b_1^1+k\widetilde b_2^2=0 \, ,
\ee
and gauge transformations 
\be\label{eq: three gauge transformations}
\begin{aligned}
A_1^1&\longmapsto A_1^1+\d\alpha \, ,
&\widetilde A_0^2&\longmapsto\widetilde A_0^2-k\alpha+2\pi N \, ,\\
b_1^1&\longmapsto b_1^1+\d\lambda_0-k\Lambda_1 \, ,
&\widetilde b_2^2&\longmapsto\widetilde b_2^2+\d\Lambda_1 \, .
\end{aligned}
\ee
The genuine topological lines are generated by
\be\label{eq: finite BF lines}
\begin{aligned}
W_n[\gamma]&=\exp\left(in\oint_\gamma A_1^1\right) \, , \\
U_\ell[\gamma']&=\exp\left(i\ell\int_{\Sigma_2}\widetilde b_2^2
+\frac{i\ell}{k}\oint_{\gamma'}b_1^1\right),
\qquad\gamma'=\partial\Sigma_2 \, ,
\end{aligned}
\ee
with $n,\ell\in\bZ$. The dressing of $U_\ell$ cancels its $\Lambda_1$ gauge variation, and the integral $2\pi$ periods of $\widetilde b_2^2$ make it independent of the choice of $\Sigma_2$. The compact scalar supplies an endpoint for $W_k$. Likewise, $U_\ell$ inserts magnetic flux $2\pi\ell/k$, so $U_k$ can terminate on a unit monopole of $A_1^1$. Thus both labels are identified modulo $k$.

The pure electric and magnetic lines are bosonic, with trivial braiding within each family, while their mutual braiding is
\be\label{eq: finite BF braiding}
\mathcal B(W_n,U_\ell)=\exp\left(\frac{2\pi i n\ell}{k}\right) \, .
\ee
The lines $W_nU_\ell$, with $n,\ell\in\bZ_{|k|}$, therefore have precisely the fusion and braiding of untwisted $\bZ_{|k|}$ gauge theory. With the Dirichlet symmetry boundary for $A_1^1$, this is the SymTFT of the finite symmetry that survives gauging $U(1)_2$. In boundary terms, the mixed anomaly becomes an ABJ anomaly for $U(1)_1$, explicitly reducing it to $\bZ_{|k|}$.

A concrete realization of a boundary theory with this symmetry is the massless charge $k$ Schwinger model. Its bosonized Euclidean action is \cite{Misumi:2019dwq}
\be\label{eq: bosonized Schwinger model}
S=\int_{X_2}\left[
\frac{1}{8\pi}\d\varphi\wedge\star\d\varphi
+\frac{1}{2g^2}f_2\wedge\star f_2
+\frac{ik}{2\pi}\varphi f_2\right] \, ,
\qquad\varphi\sim\varphi+2\pi \, ,
\ee
where $f_2=\d a_1$ is the curvature of a compact dynamical $U(1)$ connection and $g$ is the boundary gauge coupling. The $U(1)_1$ transformation is the scalar shift $\varphi\mapsto\varphi+\alpha$. In a flux sector $\frac{1}{2\pi}\int_{X_2}f_2=m\in\bZ$, this changes the action by $\Delta S=ik\alpha m$, meaning the path-integral weight is invariant in every sector precisely when $k\alpha\in2\pi\bZ$. The surviving shifts $\varphi\mapsto\varphi+2\pi\ell/k$, with $\ell\in\bZ_{|k|}$, realize the finite zero-form symmetry predicted above.\footnote{The charge-$k$ Schwinger model also has an electric $\bZ_{|k|}^{(1)}$ symmetry with a mixed anomaly with this zero-form symmetry \cite{Misumi:2019dwq}. Here we focus on the zero-form subgroup.}

\paragraph{Quantization and limits.}
Now we quantize \eqref{eq: MaxMaxCS} and derive its spectrum, from which we show that the double scaling limit followed by consistent photon truncation leaves a finite dimensional Hilbert space. The canonical analysis uses the Lorentzian formulation of \eqref{eq: MaxMaxCS}. With $E^\alpha$ normalized as in Section \ref{sec: CQ of Maxwell}, the Gauss laws read
\be\label{eq: mixed Gauss laws}
-\d\star E^1+\frac{k}{2\pi}B^2=0 \, ,\qquad
-\d\star E^2+\frac{k}{2\pi}B^1=0 \, .
\ee
On a closed surface without charged insertions they impose $km_1=km_2=0$. The coexact modes give two photon towers with frequencies \eqref{eq: mixed photon mass}. The position and momentum operators of the harmonic sector obey
\be\label{eq: mixed harmonic algebra}
[(\widehat{Q}^{\alpha})^I,(\widehat{Q}^{\beta})^J] = 0 \, ,
\qquad[(\widehat Q^\alpha)^I \, ,(\widehat P^\beta)_J]
=i\delta^{\alpha\beta}{\delta^I}_J \, , \qquad [(\widehat P^\alpha)_I,(\widehat P^\beta)_J]
=\frac{ik}{2\pi}(\sigma^1)^{\alpha\beta}\omega_{IJ},
\ee
with mutually commuting $\widehat{Q}^I$, and $\omega$ the intersection matrix on harmonic 1-forms. Their Hamiltonian is
\be\label{eq: mixed harmonic Hamiltonian}
\widehat{H}_{\mathrm h}=\sum_{I,J=1}^{2g} \left[\frac{e_1^2}{2} L^{IJ}(\widehat P^1)_I(\widehat P^1)_J
+\frac{e_2^2}{2}L^{IJ}(\widehat P^2)_I(\widehat P^2)_J \right] \, .
\ee
Removing the zero point energy of the harmonic sector, the full normal-ordered Hamiltonian reads
\be\label{eq: mixed full spectrum}
\mathopen{:}\widehat{H}\mathclose{:}=
\sum_{\alpha=1}^{2}\sum_{n \in \bN}\sqrt{\mu_n+m_{12}^2}\,
(\widehat a_n^\alpha)^\dagger\widehat a_n^\alpha
+\sum_{I,J=1}^{2g} \left[ \frac{e_1^2}{2}L^{IJ}(\widehat P^1)_I(\widehat P^1)_J
+\frac{e_2^2}{2}L^{IJ}(\widehat P^2)_I(\widehat P^2)_J \right] - g m_{12} \, .
\ee
Consider the following two sets of harmonic position and momenta operators
\be\label{eq: mixed guiding centers}
(\widehat R^1)^I=(\widehat Q^1)^I-
\frac{2\pi}{k}(\omega^{-1})^{IJ}(\widehat P^2)_J \, ,
\qquad
(\widehat R^2)^I=(\widehat Q^2)^I-
\frac{2\pi}{k}(\omega^{-1})^{IJ}(\widehat P^1)_J \, .
\ee
One can verify that such operators commute with all harmonic momenta at any finite $e_{\alpha}$, and so in particular commute with the Hamiltonian. Thus we observe that whilst $e^{i n_I (\widehat{Q}^{\alpha})^I}$ and $e^{i \alpha^I (\widehat{P}^{\beta})_I}$ create states of non-zero energy before taking the limit, there are $|k|^{4g}$ operators that preserve the zero-energy state condition
\be \label{eq: Atop generators}
    W_{\bold{n}} = e^{i n_I (\widehat{R^1})^I} \, , \qquad U_{\boldsymbol{\ell}} = e^{i \ell_I (\widehat{R^2})^I} \, , \qquad \bold{n},\boldsymbol{\ell}\in \bZ_{|k|}^{2g} \, ,
\ee
generating a $|k|^{4g}$ dimensional algebra $\cA_{\text{top}}$, for any finite $e_{\alpha}$. These operators satisfy the braiding of the $\bZ_{|k|}$ BF theory 
\be\label{eq: finite harmonic Wilson algebra}
W_{\bold n}U_{\boldsymbol{\ell}}
=e^{2\pi i n_I \ell^I/k}\,
U_{\boldsymbol{\ell}}W_{\bold n} \, .
\ee

The form of the Hamiltonian \eqref{eq: mixed full spectrum} can mislead in extracting the Hilbert space resulting from the double scaling limit since the $(\widehat{P}^1)_I$ and $(\widehat{P}^2)_I$  do not commute among themselves, and there are suitable linear combinations of them that create finite energy states even after the limit (notice how the semi-classical analysis is completely blind to this curios phenomenon). For example, on a unit square torus, normalized annihilation operators are
\be\label{eq: mixed Landau oscillators}
\widehat c_x=\frac{e_1(\widehat P^1)_x+i \, \text{sgn}(k) e_2(\widehat P^2)_y}{\sqrt{2m_{12}}} \, ,
\qquad
\widehat c_y=\frac{e_1(\widehat P^1)_y-i \, \text{sgn}(k) e_2(\widehat P^2)_x}{\sqrt{2m_{12}}} \, ,
\ee
satisfy $[\widehat c_i,\widehat c_j^\dagger]=\delta_{ij}$, and $\widehat{c}_i^\dagger$ create states of finite energy $m_{12}$. Therefore, to extract a topological theory after taking the double scaling limit, we must project out not only the photon states, but also the operators which create these non-zero energy states in the Harmonic sector. The harmonic sector separates into $2g$ oscillators and a degenerate topological sector \cite{Bertrand:2007ja}. Below we check that restricting to the oscillator vacuum preserves the topological operator algebra.

Explicitly, we keep the operators $W_{\bold{n}}$ and $U_{\boldsymbol{\ell}}$ that generate $\cA_{\text{top}}$, as well as the operators 
$V_{\bold{n}}=e^{in_I(\widehat{Q}^1)^I}$ with $\bold{n}\in \bZ^{2g}$ and $\widetilde{V}_{\boldsymbol{\alpha}}=e^{i \alpha_I (\widehat{P}^2)^I}$ with $\boldsymbol{\alpha}\in \bR^{2g}$ which commute with the Hamiltonian in the double scaling limit. However, we observe that given the set of operators $\{\widetilde{V}_{\boldsymbol{\alpha}},W_{\bold{n}},U_{\boldsymbol{\ell}} \}$, we can recover the operators $V_{\bold{n}}$. Moreover, we can argue that on the topological Hilbert space $\cH_{\text{top}}$ of normalizable states $|\psi \rangle$ with zero energy, the operators $\widetilde{V}_{\boldsymbol{\alpha}}$ act as the identity operator
\be
    ||(\widetilde{V}_{\boldsymbol{\alpha}}-1) |\psi \rangle ||^2 \le \alpha^I \alpha^J \langle \psi| (\widehat{P}^2)_I (\widehat{P}^2)_J |\psi \rangle \le \frac{g|k|}{\pi} \frac{e_1}{e_2} \alpha^I \alpha^J L_{IJ} \rightarrow 0 \, .
\ee
Thus $\cA_{\text{top}}$ is sufficient to generate all non-trivial operators after projection. We identify the operators \eqref{eq: Atop generators} with the operators $W_n$ and $U_{\ell}$ in \eqref{eq: finite BF lines} of the continuum $\bZ_{|k|}$ BF theory. The Hilbert space after projection $\cH_{\text{top}}$ is spanned by the states created by $W_{\bold{n}}$, and has dimension
\be
    \dim \cH_{\text{top}} = |k|^{2g} \, .
\ee
Thus we have recovered the $\bZ_{|k|}$ BF theory from canonical quantization.

\subsubsection{Non-Invertible Chiral Symmetry}

The next example applies the same logic to a cubic anomaly coupling. We do not perform canonical quantization of the interacting five-dimensional theory. Instead, we check the boundary gauging, identify a double scaling limit that keeps the local interaction finite, and derive the corresponding first-order topological action.  Let $A_1^1$ and $A_1^2$ couple respectively to $U(1)_A$ and $U(1)_V$, with an AVV anomaly and no other anomaly terms:
\be\label{eq: 5dMaxMaxCS}
S=\frac{1}{2e_1^2}\int_{M_5}\d A_1^1\wedge\star\d A_1^1
+\frac{1}{2e_2^2}\int_{M_5}\d A_1^2\wedge\star\d A_1^2
+\frac{ik}{8\pi^2}\int_{M_5}A_1^1\wedge\d A_1^2\wedge\d A_1^2.
\ee
Here $k\in2\bZ$. This convention preserves vector gauge invariance. We therefore impose Neumann for $A_1^2$ to gauge $U(1)_V$, while $A_1^1$ remains Dirichlet. 

After canonical normalization $A_1^\alpha=e_\alpha a_1^\alpha$, the cubic interaction has coupling proportional to $ke_1e_2^2$. We therefore consider the double scaling limit
\be\label{eq: balanced cubic limit}
e_1\rightarrow0,\qquad e_2\rightarrow\infty,\qquad
e_1e_2^2=\lambda,\qquad 0<\lambda<\infty,
\ee
that keeps the coupling finite. We postpone the full quantum analysis for the future, and here we limit ourselves to the semi-classical computation, analogous to the one we used in the (2+1)d example. To obtain the topological action, use an auxiliary real three-form $b_3^1$, an independent real two-form $f_2^2$, and a compact dual two-form $C_2^2$:
\be\label{eq: cubic first order action}
\begin{aligned}
S={}&\frac{e_1^2}{8\pi^2}\int b_3^1\wedge\star b_3^1
+\frac{1}{2e_2^2}\int f_2^2\wedge\star f_2^2\\
&+\frac{i}{2\pi}\int\left[
b_3^1\wedge\d A_1^1+f_2^2\wedge\d C_2^2
+\frac{k}{4\pi}A_1^1\wedge f_2^2\wedge f_2^2\right].
\end{aligned}
\ee
Integration over $C_2^2$ imposes that $f_2^2$ is a compact curvature. Eliminating $b_3^1$ then recovers \eqref{eq: 5dMaxMaxCS}. In the limit \eqref{eq: balanced cubic limit} both metric-dependent coefficients vanish, leaving
\be\label{eq: chiral topological action}
S_{\mathrm{top}}=\frac{i}{2\pi}\int_{M_5}\left[
b_3^1\wedge\d A_1^1+f_2^2\wedge\d C_2^2
+\frac{k}{4\pi}A_1^1\wedge f_2^2\wedge f_2^2\right].
\ee
The topological fields $b_3^1$ and $f_2^2$ are $\bR$ gauge fields with anomaly-induced transformations mixing them with the compact fields (see \cite{Antinucci:2024zjp} for the details).  With Dirichlet boundary conditions for $A_1^1$ and $C_2^2$, the latter sourcing the magnetic $U(1)^{(1)}$ symmetry of the gauged vector theory, this is the action obtained by the dynamical gauging substitution in \cite{Antinucci:2024zjp}. Here it follows from a double scaling limit of the first-order dynamical bulk action.

The cubic term explains why the axial remnant differs from the finite symmetry of the previous example. Rational axial rotations can be dressed with three-dimensional Abelian TQFTs to give genuine topological defects that lead to non-invertible fusion. This is the rational-angle chiral symmetry $\bQ/\bZ$ \cite{Choi:2022jqy,Cordova:2022ieu}, whose SymTFT is \eqref{eq: chiral topological action}. The operator construction is given in \cite{Antinucci:2024zjp}.\footnote{Extensions allowing irrational angles are discussed in \cite{Arbalestrier:2024oqg}.}

\subsection{Dynamical Descent Procedure}

In this section we contextualize our results within the SymTFT from descent framework. The case of a single $p$-form Maxwell theory has been discussed briefly in \cite{Mignosa:2025cpg}, and in the context of geometric engineering in \cite{Gagliano:2024off}. We will explain in more detail the case of a single generalized Maxwell theory, then explain how the framework generalizes to any number of $U(1)$ symmetries, including the addition of anomalies.

Consider the following $(d+2)$-dimensional bulk BF theory that is the SymTFT of $(d+1)$-dimensional $p$-form Maxwell theory
\be
    S = \frac{i}{2\pi} \int_{M_{d+2}} b_{d-p-1} \wedge \d \widetilde{b}_{p+2} \, ,
\ee
involving two $\bR$-gauge fields. The theory has topological operators
\be
     U_{x}[\Sigma_{d-p-1}] = \exp \left( i x \oint_{\Sigma_{d-p-1}} b_{d-p-1} \right) \, , \qquad V_{y}[\Sigma_{p+2}] = \exp \left( iy \oint_{\Sigma_{p+2}} \widetilde{b}_{p+2} \right) \, ,
\ee
with $x,y\in \bR$, and braiding
\be
    U_x[\Sigma_{d-p-1}] V_y[\Sigma_{p+2}] = e^{2\pi i xy \text{Link}(\Sigma_{d-p-1},\Sigma_{p+2})} V_y[\Sigma_{p+2}] U_x[\Sigma_{d-p-1}] \, .
\ee
Notice that on a manifold $M_{d+2}$ with boundary $\partial M_{d+2}=X_{d+1}$, this action is not gauge invariant under
\be
    b_{d-p-1} \rightarrow b_{d-p-1}+\d \lambda_{d-p-2} \, , \qquad \widetilde{b}_{p+2} \rightarrow \widetilde{b}_{p+2} + \d \widetilde{\lambda}_{p+1} \, ,
\ee
since the action acquires the boundary variation
\be
    \delta S = \frac{i}{2\pi} \int_{X_{d+1}} \lambda_{d-p-2} \wedge \d \widetilde{b}_{p+2} \, .
\ee
To cancel this, we add the boundary action
\be
    S_{\partial} = (-1)^{d-p-1}\frac{i}{2\pi}\int_{X_{d+1}} (b_{d-p-1} \wedge \d A_{p+1}-b_{d-p-1}\wedge \widetilde{b}_{p+2}) \, ,
\ee
with gauge variation $\delta A_{p+1}=\widetilde{\lambda}_{p+1}$ such that $\delta S+\delta S_{\partial}=0$. The boundary value of the field $\widetilde{b}_{p+2}|_{X_{d+1}}$ is interpreted as the background gauge field of a $(p+1)$-form symmetry. Turning off the background gauge field $\widetilde{b}_{p+2}$, we recognize the action $S_{\partial}$ as the SymTFT of a $U(1)^{(p)}$. 

There is an analogous story in which one starts with the SymTFT with Lagrangian $\widetilde{b}_{p+2}\wedge \d b_{d-p-1}$ on $M_{d+2}$, in which case the boundary action required for canceling the boundary gauge variation is
\be
    S_{\partial} = (-1)^{p}\frac{i}{2\pi} \int_{X_{d+1}} (\widetilde{b}_{p+2}\wedge \d \widetilde{A}_{d-p-2}-\widetilde{b}_{p+2} \wedge b_{d-p-1} ) \, .
\ee
In this case, the gauge variation is $\delta \widetilde{A}_{d-p-2}=\lambda_{d-p-2}$, and $b_{d-p-1}|_{X_{d+1}}$ is a background gauge field for the $U(1)^{(d-p-2)}$ symmetry on the boundary Maxwell theory. Turning off the background field $b_{d-p-1}$, we recognize the action $S_{\partial}$ as the SymTFT of a $U(1)^{(d-p-3)}$. 

We now construct a family of gapless boundary conditions, given by $p$-form Maxwell theory. Consider first the boundary condition defined by terminating the operators $V_{nx_0}$ and $U_{\frac{n}{x_0}}$ for $n \in \bZ$ and $x_0 \in \bR^{\times}$. Then on the boundary we get a $ U(1)^{(p+1)} \times U(1)^{(d-p-2)}$ symmetry. The operators $V_{\alpha x_0}$ and $U_{\frac{\beta}{x_0}}$ for $\alpha \in \bR/\bZ \cong U(1)$ and $\beta \in \bR/\bZ\cong U(1)$ are confined on this boundary. 

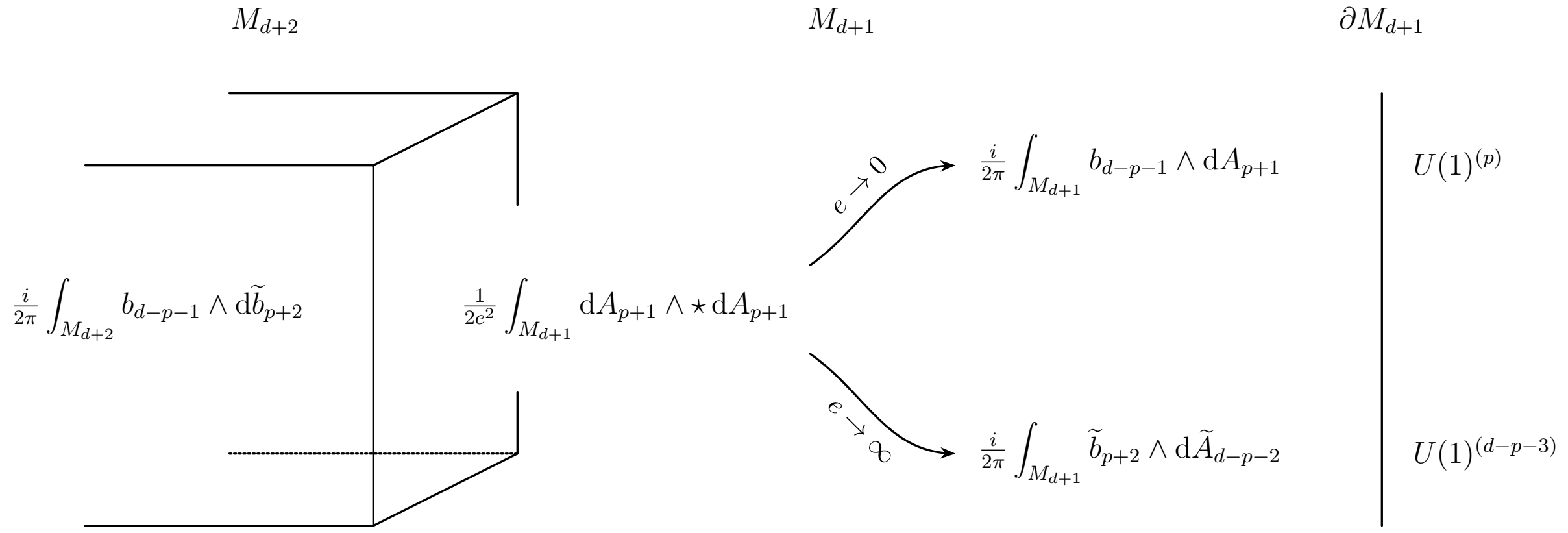
\begin{figure}[htbp]
    \centering
    \resizebox{\linewidth}{!}{%
    \begin{tikzpicture}[
        font=\large,
        line cap=round,
        line join=round,
        surface/.style={
            draw=black,
            line width=0.85pt
        },
        action/.style={
            inner sep=5pt,
            outer sep=0pt,
            fill=white
        },
        flow/.style={
            draw=black,
            line width=0.85pt,
            -{Stealth[length=6pt,width=4.5pt]}
        },
        limit/.style={
            font=\normalsize,
            fill=white,
            inner sep=2pt
        }
    ]

    \draw[surface] (-4,0) -- (0,0) -- (0,5) -- (-4,5);
    \draw[surface] (0,5) -- (2,6) -- (-2,6);
    \draw[surface] (0,0) -- (2,1);
    \draw[surface, densely dotted] (-2,1) -- (2,1);
    \draw[surface] (2,6) -- (2,4.45);
    \draw[surface] (2,1) -- (2,1.85);

    \node[action] (bulk) at (-3,3) {%
        $\displaystyle
        \tfrac{i}{2\pi}
        \int_{M_{d+2}}
        b_{d-p-1}\wedge \mathrm{d}\widetilde{b}_{p+2}$
    };

    \node[action] (maxwell) at (3.5,3) {%
        $\displaystyle
        \tfrac{1}{2e^2}
        \int_{M_{d+1}}
        \mathrm{d}A_{p+1}\wedge
        \star\,\mathrm{d}A_{p+1}$
    };

    \node[action] (electric) at (10.5,5) {%
        $\displaystyle
        \tfrac{i}{2\pi}
        \int_{M_{d+1}}
        b_{d-p-1}\wedge \mathrm{d}A_{p+1}$
    };

    \node[action] (magnetic) at (10.5,1) {%
        $\displaystyle
        \tfrac{i}{2\pi}
        \int_{M_{d+1}}
        \widetilde{b}_{p+2}\wedge
        \mathrm{d}\widetilde{A}_{d-p-2}$
    };

    \draw[flow]
        ([xshift=3pt]maxwell.north east)
        to[out=35,in=180]
        node[limit, pos=0.48, above=4pt, sloped]
            {$e\to 0$}
        ([xshift=-4pt]electric.west);

    \draw[flow]
        ([xshift=3pt]maxwell.south east)
        to[out=-35,in=180]
        node[limit, pos=0.48, below=4pt, sloped]
            {$e\to\infty$}
        ([xshift=-4pt]magnetic.west);

    \draw[surface] (14,0) -- (14,6);

    \node[anchor=west] at (14.3,5)
        {$U(1)^{(p)}$};
    \node[anchor=west] at (14.3,1)
        {$U(1)^{(d-p-3)}$};

    \node at (-1.5,7) {$M_{d+2}$};
    \node at (6.5,7)  {$M_{d+1}$};
    \node at (14,7)   {$\partial M_{d+1}$};

    \end{tikzpicture}%
    }
    \caption{Dynamical descent procedure from the
    $(d+2)$-dimensional SymTFT. $U(1)^{(p)}$ and
    $U(1)^{(d-p-3)}$ denote the global symmetries realized
    on the boundary $\partial M_{d+1}$ in the two Maxwell limits.}
    \label{fig:dynamical descent symtft}
\end{figure}

Let us match this boundary condition with extended Maxwell operators. Observe that the operators $V_{nx_0}[\Sigma_{p+2}]$ endpoints are Maxwell Wilson operators 
\be
    \exp \left( in \oint_{\partial \Sigma_{p+2}} A_{p+1} \right) \, .
\ee
Moreover, the endpoints of operators $U_{\frac{n}{x_0}}[\Sigma_{d-p-1}]$ are 't Hooft operators of generalized Maxwell theory
\be
    \exp \left( in \oint_{\partial \Sigma_{d-p-1}} \widetilde{A}_{d-p-2} \right) \, .
\ee
The operators $V_{\alpha x_0}$ and $U_{\frac{\beta}{x_0}}$ are confined on this boundary and give the magnetic and electric Gukov Witten operators of Maxwell theory
\be
    \exp \left( i \frac{\beta}{x_0} \oint_{\Sigma_{p+2}} \widetilde{b}_{p+2} \right) \, , \qquad\exp \left( i \alpha x_0 \oint_{\Sigma_{d-p-1}} b_{d-p-1} \right) \, .
\ee
Interestingly, this gives us a family of gapless boundaries labelled by a parameter $x_0$ that we can interpret as the coupling $e$ of the Maxwell theory we put on the boundary. See Figure \ref{fig:dynamical descent symtft} for a visual representation of this setup.

Tuning this parameter allows us to interpolate between the two different $e=0$ and $e=\infty$ boundary conditions via a family of Maxwell theories labelled by the value of the coupling $e$. In the limit $e \rightarrow 0$, the 't Hooft operators are decoupled from the spectrum, meaning there are no operators for $U_{\frac{n}{x_0}}[\Sigma_{d-p-1}]$ to end on, and thus these operators are confined to the boundary. From the SymTFT$_{d+2}$ perspective, all of the topological operators $U_{x}[\Sigma_{d-p-1}]$ are confined on the boundary, generating a trivially acting $\bR$ symmetry. Thus the Maxwell theory in this limit only has a $U(1)^{(p+1)}$ symmetry coming from the operators $V_y[\Sigma_{p+2}]$. Since the Wilson lines become topological in the $e\rightarrow 0$ limit, this explains why this limit is the SymTFT for a $U(1)^{(p)}$ symmetry. The $e\rightarrow \infty$ limit works analogously. 

The framework for building the dynamical descent picture for an arbitrary number of $U(1)$ symmetries is therefore a clear generalization. For each $U(1)^{(p)}$ symmetry in the $d$-dimensional theory, we take a bulk $p$-form Maxwell theory, which can be recognized as the gapless boundary of a $(d+2)$-dimensional SymTFT. Additionally, for each anomaly in the $d$-dimensional theory, we may add to the Maxwell theory in $(d+1)$-dimensions the appropriate anomaly inflow action, and to the SymTFT in $(d+2)$-dimensions add the anomaly polynomial. 

Building the SymTFT for the $d$-dimensional theory is then a case of taking the $e \rightarrow 0$ limit of the gapless SymTFT$_{d+2}$ boundary theory for each Maxwell factor. If one wanted instead to first gauge a $U(1)$ symmetry in the $d$-dimensional theory, they would instead take the $e \rightarrow \infty$ limit of the SymTFT$_{d+2}$ boundary for the relevant $U(1)$ factor.

\subsection*{Acknowledgments}
\noindent
We are grateful to Daniel Brennan, Inaki Garcia-Extebarria, and Finn Gagliano for discussions, and in particular to Sakura Schafer-Nameki for collaboration on a related project and feedback on the draft. 
Some of the computations were done and checked with the help of ChatGPT/Codex Sol 5.6 and Astra 6. 
The work of AA is supported by the UKRI Frontier Research Grant, underwriting the ERC Advanced Grant ``Generalized Symmetries in Quantum Field Theory and Quantum Gravity''. OL is supported by an STFC studentship.

\small{ 
\bibliographystyle{ytphys}
\baselineskip=0.85\baselineskip
\bibliography{GenSym}
}

\end{document}